\documentclass{aa}  

\usepackage{graphicx}
\usepackage{natbib}
\newcommand{\SII}{[SII]$\lambda\lambda$6716,6731\AA}	
\newcommand{\NII}{[NII]$\lambda\lambda$6548,6584\AA}	
\newcommand{\Ha}{H$\alpha \lambda$6563\AA}	
\newcommand{\OI}{[OI]$\lambda\lambda$6300,6364\AA}	
\usepackage{hyperref}
\begin{document} 

   \title{Optical spectroscopy of type-2 LINERs}

   \author{L. Hermosa Mu\~{n}oz \inst{1}
        \and
        S. Cazzoli\inst{1}
        \and
        I. M{\'a}rquez\inst{1}
        \and
        J. Masegosa\inst{1}
        }

   \institute{
        1. Instituto de Astrof\'isica de Andaluc\'ia - CSIC, Glorieta de la Astronom\'ia s/n, 18008 Granada, Spain \\
              \email{lhermosa@iaa.es}
             }

   \date{Received M DD, YYYY; accepted M DD, YYYY}

 
  \abstract{
  Type-2 Low-ionization Narrow Emission-line Regions (LINERs) have been optically classified with the Palomar data as not presenting a broad component in the Balmer emission lines which is associated to the Broad Line Region (BLR) of the Active Galactic Nuclei (AGN).}
  {We aim to unveil the presence of different kinematic components of emission lines in the nuclear region of a sample of local (z\,$\leq$\,0.022) type-2 LINERs. We focus on the analysis of the true nature of LINERs by means of the detection (or not) of a broad component originated in the BLR of the AGN. Additionally, we search for the possible presence of non-rotational motions such as outflows in these LINERs.}
  {We have applied a decomposition of the nuclear emission lines using an  spectroscopic analysis of the optical spectra of 9 type-2 LINERs of intermediate-resolution spectroscopic data retrieved from the Hubble Space Telescope (\textit{HST}) archive. The study is completed with archival spectra from the Double Spectrograph from the Palomar Observatory. }
  {The emission line fitting reveals the presence of a broad component associated to the BLR in 6 out of the 9 galaxies for the space-based data, and for 2 out of the 8 from the ground-based spectra. The velocity dispersion for two galaxies (NGC 4486 and NGC 4594) measured in \textit{HST}/STIS data suggest the presence of outflows.}
  {The results indicate that the spatial resolution plays a major role in the detection of the BLR, as it appears diluted in the ground-based data (even after removing stellar contribution). This is also true for the emission line diagnostics, as the contaminant light contributes to lower emission line ratios toward the star-forming area of standard BPTs. 
  We propose to reclassify NGC 4594 as a type-1 LINER, since a BLR component is seen in both space- and ground-based spectra. We find ambiguous results for the BLR component of NGC 4486. The modest outflow detection in our sample may indicate that they are not be as frequent as seen for type-1 LINERs.}


   \maketitle
%

 \section{Introduction}

   Low-ionization Narrow Emission-line Regions (LINERs) are among the lowest-luminosity active galactic nuclei (AGN) found in the Universe. Defined for the first time by \citet{Heckman1980}, their spectra is dominated by low-ionization lines in contrast to the importance of high-ionization emission lines ([O\,III]$\lambda 5007$ or He\,II$\lambda 4686$) in more luminous AGNs, as Seyferts. The initial separation between LINERs and Seyferts was defined by \citet{Heckman1980} based on the relative ratios between the intensity of several emission lines ([O\,II]$\lambda 3727$ to [O\,III]$\lambda 5007$ and [O\,I]$\lambda 6300$ to [O\,III]$\lambda 5007$).
   
   LINERs may be the most numerous type of AGNs in the local Universe \citep{Ho1997}, although their true AGN nature is still a matter of debate \citep{Ho2008, Marquez2017}. Their spectral features cannot be explained only with stellar formation processes \citep{Heckman1980}, but an AGN is not the only valid explanation. Other mechanisms that could reproduce their spectral features are, for example, shock ionization \citep{Heckman1980,Dopita1995,Dopita1996} and photoionization by post-AGB stars \citep{Binette1994,Stasinska2008,Papaderos2013}.
   
   The morphology of the circumnuclear H$\alpha$ emission as seen by the \textit{HST} imaging indicates the possible presence of outflowing material in some of the systems \citep{Pogge2000,Masegosa2011}.
   In these cases, the outflow could contribute to the broadening of emission lines such as H$\alpha$-[N\,II], [O\,I] and [S\,II]. This could compromise the detection of the weak emission of Balmer lines from the Broad Line Region (BLR) in their nucleus. Shocks associated to outflows may be important also in ionizing the nuclear (and its surroundings) gas \citep{Dopita1996, Molina2018}. However, these are observed at lower velocities \citep{Molina2018} than those needed by the shocks models \citep{Groves2004} to reproduce LINER-like line ratios.
   
   As Seyfert galaxies, LINERs appear as type-1.9 or type-2, depending on if the BLR is detected on their nucleus. The BLR is identified as a very broad component in the profiles of the Balmer lines. There are only a few systematic works studying the AGN nature of LINERs \citep{Ho2003,Cazzoli2018}, being most of them individual discoveries \citep{Bower1996, Ho1997}. 
      
   In the work by \citet{Cazzoli2018}, the spectra for a sample of nearby (z $<$ 0.025) type-1.9 LINERs obtained with both ground- and space-based telescopes are analyzed. 
   A broad component indicative of the BLR is required for the fitting of ground-based data in only a third of the targets, despite they where all classified as type-1.9 by \citet{Ho1997}. Nevertheless, the modeling of the spectra of the targets obtained with the Space Telescope Imaging Spectrograph (STIS), mounted on the Hubble Space Telescope \textit{(HST)}, require a broad H$\alpha$ feature.  
   
   As in type-2 LINERs the BLR is not detected, the AGN nature of these systems is still controversial. Other components, as post-AGB stars or low mass X-ray binaries may play an important role to explain their optical spectra \citep[see ][]{Ho1997, McKernan2010}. 
   
   In this paper we analyze \textit{HST} spectra of 9 type-2 LINERs selected from the sample of \citet{GM2009} with available data in the archive to investigate if they present a broad component in the Balmer lines that could be associated to the existence of a BLR in their nuclei. We have completed the study by comparing with the ground-based spectra from \citet{Ho1995}, since the majority of the targets have been classified as type-2 objects on that work. We also revisited the possible influence of the outflows in the broadening of Balmer lines, following \citet{Cazzoli2018}.
   
   This paper is organized as follows. 
   In Section~\ref{Sample_data} we describe the sample and the data reduction process. Section~\ref{analysis} indicates the spectral line fitting for each nuclear spectra. Section~\ref{main_HST_results} and Section~\ref{main_palomar_results} shows the main modeling results for space- and ground-based spectra, respectively. In Section~\ref{discussion} we discuss the modeling results, the possible presence of a BLR in some of the targets, and a comparison of the two data-sets for both kinematics and line ratios. Finally, in Section~\ref{summary_conclusions} we present a summary and the main conclusions of the work.
   
\section{Sample and data processing}
\label{Sample_data}

   The sample selection is drawn from the survey by \citet{GM2009} of 82 LINERs selected at X-rays.
   From this sample, we selected all type-2 LINERs classified as AGNs from the analysis of their features in the wavelength range from X-rays to the near-infrared. 
   The final sample of 12 galaxies comprises all the objects with available spectra in the \textit{HST}/STIS archive with the intermediate resolution grating G750M (R $\sim$ 5000).
   
   This grating covers a total wavelength range from 5450\,-\,10140\AA, centered at 6581 or 6768 \AA. This range has been chosen due to the availability of the \SII; \NII; and \Ha\, lines in all the spectra and \OI\, in those where the grating was centered at 6581\AA\,(all except NGC 4594 and NGC 4676B). 
   Two out of 12 were discarded because the low signal-to-noise ratio (S\,/\,N) of the spectra (NGC 4261 and NGC 5055, for which \citealt{Constantin2015} already reported the poor quality of the nuclear spectrum of the latter), and also NGC 6240, because even though its spectrum is archived as to be nuclear, the emission lines were not visible. This leaves with a total of 9 galaxies in the final sample of type-2 LINERs. Their basic information is provided in Table~\ref{sample}. 

   The full calibrated 2D spectra were retrieved from the \textit{HST} archive. In Table~\ref{obslog} the main observing details of each nuclear spectrum analyzed are listed. 
   Then we extracted the 1D flux and wavelength\,-calibrated nuclear spectra of the sources. Some of them (namely NGC 4698, NGC 4552, NGC 4676B) showed abundant cosmic rays (CR). 
   The results can be improved by performing a new CR correction with the \textit{L.A. Cosmic} algorithm \citep{vanDokkum2001} to the flat-fielded frames downloaded also from the archive. After this correction, we applied the last step of the reduction process following the \textit{HST} pipeline, which is the \textsc{x2d} task (under the \textsc{stsdas} package in \textsc{IRAF}\footnote{IRAF is the Image Reduction and Analysis Facility distributed by the National Optical Astronomy Observatories (NOAO) for the reduction 
   and analysis of astronomical data. \url{http://iraf.noao.edu/}}). This task generates the final calibrated frame and corrects for the geometrical distortion. 
   
   For the galaxies with several nuclear exposures available, they were realigned and combined with the \textsc{IRAF} tasks \textsc{imshift} and \textsc{imcombine}. These tasks improved the S/N after the CR rejection. If there was only one nuclear spectrum available and the CRs fell close to the lines in the majority of the nuclear rows, we extracted each row individually and did the CR removal manually. Then the individual rows were combined to form the final nuclear spectra. If a CR fell on the continuum of these spectra, it was masked in the analysis.
   
   The final nuclear spectrum was extracted using \textsc{apall} task in \textsc{IRAF}. The number of extracted and combined rows depends on the slit width and the binning of the CCD. 
   The majority of the targets were measured with a slit width of 52$\arcsec$\,$\times$\,0.2$\arcsec$, a plate scale of 0.05$\arcsec$\,pixel$^{-1}$ and no binning. For these, the 5 central rows (pixels) were extracted to form the final spectra. For the spectra binned to 0.1$\arcsec$pixel$^{-1}$, the 3 central rows were extracted. All the galaxies observed with a slit width of 52$\arcsec$\,$\times$\,0.1$\arcsec$ were unbinned, and therefore we extracted 5 rows. 
   The number of pixels extracted for each galaxy and its equivalent spatial scale are indicated in Table~\ref{obslog} (column 6).
   
   However the data for two of the objects (NGC 4552 and NGC 4676B) had low S/N that did not improve with the combination of multiple exposures. As for NGC 4552 there were additional spectra from regions near the very center, a line-fitting was possible for the nuclear spectrum (see Sect.~\ref{analysis}). We also fitted the NGC 4676B spectrum, but it was finally excluded due to the low significance of the obtained modeling, as will be discussed in Section~\ref{discussion}.
   
   Additionally, we searched for the spectra of the 9 objects within the sample by \citet{Ho1995} in order to compare the space-based and ground-based spectra. These latter were retrieved from the archival. They were observed with the Double Spectrograph \citep{Oke1982}, located in the Cassegrain focus of the Hale 5 meter telescope at Palomar Observatory during 1984 and 1985. The long-slit spectra were measured with a 1200\,line\,mm$^{-1}$ grating blazed to cover a wavelength range $\sim 6210 - 6860$\AA\,. The spectral resolution was R $\sim$ 2500 for a 2$\arcsec$-length slit and R $\sim$ 4000 for a 1$\arcsec$-length slit. Spectra for 8 out of the 9 galaxies (all except NGC 4676B, member of a merging system, which is located at approximately 10 times the distance of the rest of the targets, see Table~\ref{sample}) were retrieved from the NASA/IPAC Extragalactic Database (NED)\footnote{The NASA/IPAC Extragalactic Database (NED) is funded by the National Aeronautics and Space Administration and operated by the California Institute of Technology.} archive already reduced. 
   The width of the slit was 2$\arcsec\times$128$\arcsec$ for all the targets except for NGC 4486 that was 1$\arcsec \times$128$\arcsec$. The total extraction window was 2$\arcsec \times$4$\arcsec$ (for NGC 4486 was 1$\arcsec \times$4$\arcsec$), equivalent to summing up the central 7 pixels of the CCD. The mean seeing of the observations is 1.5$\arcsec$. The spectral and spatial resolution was $\sim$2.5\AA\, and 0.58$\arcsec$\,pix$^{-1}$, respectively.

\begin{table*}
  \caption{General properties for the 9 type-2 LINERs discussed in this paper. (2) \lq RA\rq and (3) \lq DEC\rq: coordinates of the galaxy; (4) \lq Morphology\rq: Hubble classification; (5) \lq z\rq and (6) \lq scale\rq: heliocentric redshifts and scale distance from the Local Group from NED; (7) \lq \textit{i}\rq: inclination angle from \citet{Ho1997}; (8) \lq V$_{\rm rot}$\rq\, and (9) \lq P.A.\rq: respectively, maximum rotation velocity corrected from inclination and position angle from HyperLeda.} 
  \centering
  \begin{tabular}{l c c c c c c c c}
    \hline \hline
    ID & RA & DEC & Morphology & z & Scale & \textit{i} & V$_{\rm rot}$ & P.A. \\
    & (hh:mm:ss) & (dd:mm:ss) &  &  & (pc\,arcsec$^{-1}$) & ($^{\circ}$) & (km\,s$^{-1}$) & ($^{\circ}$) \\ 
    (1) & (2) & (3) & (4) & (5) & (6) & (7) & (8) & (9) \\
   \hline
    NGC 2685 & 08 55 34.71 & +58 44 03.83 & (R)SB0+pec$^{(a)}$ & 0.0029 &  70 &  60 & 146$\pm$4 & 38 \\
    NGC 3245 & 10 27 18.387 & +28 30 26.79 & SA(r)0$^{(b)}$ &  0.0045 &  90 & 58 & - & 177 \\
    NGC 4374 & 12 25 03.74 & +12 53 13.14 &  E1$^{(c)}$  & 0.0034 &  65 & - & 189$\pm$15 & 135 \\
    NGC 4486 & 12 30 49.42 & 12 23 28.04 & E0+pec$^{(d)}$ & 0.0043 & 84 &  - & - & 160 \\
    NGC 4552 & 12 35 39.81 & +12 33 22.83 & E$^{(a)}$ & 0.0011 &  17 & - & - & - \\                 
    NGC 4594 & 12 39 59.43 & -11 37 22.99 & SA(s)a$^{(a)}$ & 0.0034 &  59 &  68 & 408$\pm$11 & 90 \\
    NGC 4676B & 12 46 11.24 & +30 43 21.87 & SB(s)0/a+pec$^{(a)}$ & 0.0220 & 472 & - & 397$\pm$31 & - \\
    NGC 4698 & 12 48 22.91 & +08 29 14.58 & SA(s)ab$^{(a)}$ & 0.0034 &  64 & 53 & 518 & 203$\pm$3 \\
    NGC 4736 & 12 50 53.061 & +41 07 13.65 & (R)SA(r)ab$^{(a)}$ & 0.0010 &  25 & 36 & 182$\pm$5 & 105 \\               
    \hline
  \end{tabular}
\label{sample}
\tablebib{$^{(a)}$\citet{Ho1997}; $^{(b)}$\citet{Wardle1986}; $^{(c)}$\citet{Huchtmeier1994}; $^{(d)}$\citet{Huchtmeier1986}.}
\end{table*}

\begin{table*}
	\caption{Observing log of \textit{HST}/STIS data. Columns indicate: (1) galaxy name; (2) other name; (3) filename of the analyzed spectra as indicated in the archive; (4) proposal ID of the observations; (5) position angle of the slit (north-eastward); (6) extraction window and its corresponding scale in parsecs; (7) date of the observation; (8) exposure time of the spectra; (9) filter used for the sharp-divided image \citep{Marquez1996} obtained also from the \textit{HST} archive. These latter are shown in Appendix~B.}
	\label{obslog}
    \centering
	\begin{tabular}{l c c c c c c c c}
	\hline \hline
	ID & Other name & Filename & Proposal & PA & Extr. window & Obs. date & Exp.Time & Im.Filter\\ 
	 & & (\_sx2\,/\_x2d) & ID & ($^{\circ}$) & pix (pc) & (yy-mm-dd) & (s) & \\
	 (1) & (2) & (3) & (4) & (5) & (6) & (7) & (8) & (9) \\ \hline
	NGC2685* & IRAS\,08517+5855 & o63n01020 & $8607$ & 54 & 3 (21) & 2001-05-06 & $3097$ & F814W \\
	NGC3245 & IRAS\,10245+2845 & o57205030 & $7403$ & 203 & 5 (23) & 1999-02-02 & $2715$ & F547M \\
	NGC4374 & M\,84 & o3wn01010/2010 & $7124$ & 104 & 5 (17) & 1997-04-14 & $1993/2223$ & F814W \\		
	NGC4486 & M\,87 & o67z01010/2010 & $8666$ & 164 & 5 (21) & 2001-03-25/27 & $1360$ & F606W \\		
	NGC4552 & M\,89 & o5l203020 & $8472$ & 78 & 5 (4) & 2000-06-19 & $1440$ & F555W \\	
	NGC4594 & M\,104 & o4d303030 & $7354$ & 250 & 5 (15) & 1999-02-05 & $280$ & F814W \\	
	NGC4676B* & IC\,820 & o67q09030 & $8669$ & 195 & 3 (144) & 2002-02-20 & $740$ & F814W \\	
	NGC4698 & IRAS\,12458+0845 & o4e022010/20/30 & $7361$ & 259 & 5 (16) & 1997-11-24 & $900/933/840$ & F606W \\
	NGC4736 & M\,94 & o67110030 & $8591$ & 50 & 5 (6) & 2002-07-15 & $1440$ & F555W \\	 \hline
	\end{tabular}
\tablefoot{*: spectrum obtained with a slit width whose plate scale was 0.102$^{\arcsec}$pix$^{-1}$. It is 0.051$^{\arcsec}$pix$^{-1}$ for the rest of the objects.}
\end{table*}

\section{Analysis of the nuclear spectra}
\label{analysis}

   The host galaxy contribution on the nuclear spectra of a low-luminosity AGN may be significant, hence the starlight should be subtracted to robustly measure the kinematics and fluxes of the emission lines. 
   However, due to the small wavelength range covered for the \textit{HST}/STIS data (572\,\AA), we decided not to perform the subtraction as the line free continuum regions of the spectra  are small for a proper stellar continuum modeling.
   Moreover, \citet{Constantin2015} (hereafter C15) proved that the correction is negligible for \textit{HST}/STIS spectra in this wavelength range due to the small aperture of the instrument, that reduces the contaminant starlight from the host galaxy. 
   
   The starlight was modeled and subtracted for the Palomar data using a penalized PiXel fitting analysis \citep[\textsc{ppxf} version 4.71][]{Cappellari2004,Cappellari2017}. This differs to the stellar modeling applied by \citet{Ho1993}, in which template galaxies were used instead (see Sect.~\ref{comparison_Ho}). 
   The methodology of the stellar subtraction was done following a similar technique as in \citet{Cazzoli2018} (hereafter C18). Similarly for the modeling of the spectral emission lines, which is explained in below. For each object, the restframe velocity used in the analysis is defined as: \textit{c\,$\times$\,z} (column 5, Table~\ref{sample}).
   The starlight decontamination for the Palomar data and all the line modelings to the different spectra are found in Appendices~\ref{appendix_B} and~\ref{appendix}, respectively. 
   
\subsection{Emission line fitting}
\label{analysis_LF}
   
   Each of the emission lines present in the spectra ([S\,II], [N\,II], H$\alpha$ and in some cases [O\,I], Table~\ref{T_kin} column 4) were fitted with Gaussian functions to derive their properties: central wavelength, line width ($\sigma$) and full-width half maximum (FWHM), flux and f$_{\rm blend}$, defined as the percentage of the broad component flux with respect to the total flux of the H$\alpha$-[N\,II] lines. The fitting was performed using a non-linear least-squares minimization and curve-fitting routine (\textsc{lmfit}) implemented in \textsc{Python}, which is an extension of the Levenberg-Marquardt method found in the \textsc{scipy} package. 
   
   The modeling of the lines was performed with three different methods using either [SII] or [OI] or both as templates for the H$\alpha$-[N\,II] complex (\citealt{Cazzoli2018} and references therein). As H$\alpha$ and [N\,II] lines are generally blended, we do not use them directly because the overall fit could be affected. Among those emission lines available, the best lines are the [O\,I] lines as they are separated enough between them to be good templates. However, these lines are not always available or have low S/N (e.g. see Fig.~\ref{Panel_NGC4552}), so we have to rely on [S\,II] lines, as done in previous works \citep{Ho1997,Balmaverde2014,Constantin2015}. Therefore, the three different models we can apply are the use of [S\,II], [O\,I] or both (if available) as the reference lines because they are usually not strongly blended as H$\alpha$-[N\,II]. 
   
   Following \citet{Cazzoli2018}, the first method is called S-method, and consists on fitting the [S\,II] lines and then tie to them the central wavelengths and widths of all the other narrow lines. 
   The second method (O-method) is the same as S-method but using as reference the [O\,I] lines. The third model is a mixed-model (M-model), that uses simultaneously both [S\,II] and [O\,I] lines as a template for the H$\alpha$-[N\,II] blend. 

  We tested different methods of fitting since the forbidden lines ions of the spectra have different critical densities ([S\,II] $\sim 10^{3}$\,cm$^{-3}$, [O\,I] $\sim 10^{6}$\,cm$^{-3}$ and [N\,II] $\sim 10^{4}$\,cm$^{-3}$, respectively) caused by the possible stratification in the NLR. Therefore the profiles of each line could be different (see \citealt{Balmaverde2014}).

   The fluxes of the two [N\,II] lines and the two [O\,I] lines were set to follow the relations 1:3 and 1:2.96, respectively \citep{Osterbrock2006}. For more details on the methods and the constraints, we refer to C18.     
   
   As in C18 in order to prevent over fitting we calculated a parameter, $\varepsilon_{\rm c}$, which is the standard deviation of a region of the continuum that did not have any absorption or emission line. This parameter was always calculated in a 50\AA\,-length region of the spectra. We compared it with the standard deviation of the residuals under each emission line and if $\varepsilon^{\rm line}$\,$<$\,3\,$\times$\,$\varepsilon_{\rm c}$, then the fit is considered to be reliable. For the cases where $\varepsilon_{\rm c}$ was not sufficient in order to differentiate between two different fits, we used the $\chi^{2}$ of the fit and/or visual inspection as the last criteria to select the final model.
   
   The procedure was organized in three steps.
   First, we fitted the continuum to a linear least-squares model using the regions between the emission lines visible in the spectra.
   
   Second, all lines were fitted to a single Gaussian profile (hereafter \lq narrow component\rq) with the same parameters for all of them. If the fit was satisfactory for the reference lines ([S\,II] or [O\,I]) but not for [N\,II] and H$\alpha$, then a broad Gaussian profile was added to H$\alpha$ (hereafter \lq broad component\rq). However, if the fit was not reliable in any of the lines or if the narrow component had a large velocity ($\geq$ 300\,km\,s$^{-1}$), then a second component was added to all the lines (hereafter \lq secondary component\rq).    
   
   Third, if the H$\alpha$-[N\,II] blend is not well reproduced (significant residuals) with a narrow plus a secondary component, then we added a broad component in H$\alpha$.
   All the line modelings are presented in the figures of the Appendix~\ref{appendix}.
   
   This method was not sufficient for the most complicated cases, for which additional restrictions were required. For example, the modeling of the reference lines for the NGC 4594 \textit{HST}/STIS spectrum was good enough, but the maximum of the profiles of [N\,II] and H$\alpha$ seemed to be shifted ($\sim$1\,-\,3\,\AA) from the maximum of the Gaussian profiles. Thus we shifted 1\AA\,the maximum of the secondary Gaussian component to improve the fit. This produced a change in the velocity compatible within the initial error estimation.
   
   For the \textit{HST}/STIS spectra of two galaxies (namely NGC 4374 and NGC 4552) the [S\,II] lines were so blended and the [O\,I] lines almost absent, that it was challenging to perform the modeling using them as reference (Fig.~\ref{Panel_NGC4374} and~\ref{Panel_NGC4552}). Thus we decided to fit the spectrum of a region situated a few pixels away from the nucleus and use this result to model the narrow component of the nuclear region. This was motivated by the assumption that the narrow component is tracing the kinematics of the disc of the galaxy, thus it should also be visible in regions outside the very center of the system. Moreover, by moving to an outer part of the galaxy, the contribution of the BLR (if present) should disappear due to its unresolved nature at \textit{HST} scale. As the disc should be rotating with the same velocity dispersion, then a modeling of this component in the spectra of an outer region would give approximately the same contribution that we should expect in the nucleus, but with a shift in velocity. 
   We note though that for some galaxies this condition is not completely true, as the disk velocity dispersion was found to increase towards the center \citep[for NGC 3245 and NGC 4594 see][ respectively]{Barth2001,Emsellem2000}. This is discussed in more detail for each individual case in Sect.~\ref{ind_com_modelling_BLR}.
   
   An additional assumption was made for NGC 4374. If the narrow component was fixed as within the previous method, when proceeding with the rest of the modeling the secondary component was broad enough to be compatible with a very broad AGN component of $\sigma >$ 800 km/s. The addition of a third Gaussian in the H$\alpha$-[N\,II] complex generated instead a narrow profile to fit a peak that is visible in the spectrum (see Fig.~\ref{Panel_NGC4374}). Thus we exchanged the secondary and broad components to improve the fit. 
   
   We fixed the velocity dispersion of the narrow component for NGC 4374 and NGC 4552 with this method, and allowed the velocity to vary in the nuclear spectra. This improved the final residuals and visually with respect to the normal fitting process. 
   
   The velocities, V, and line widths, $\sigma_{\rm line}$, obtained for each galaxy, as well as the number of components and the model used to fit the reference lines are indicated in Table~\ref{T_kin}. The S-method was used for all cases except for the Palomar spectrum of NGC 4552, for which we used the O-method. The line width has been corrected from the instrumental width by doing $\sigma_{\rm line} = \sqrt{\sigma_{\rm obs}^{2} - \sigma_{\rm inst}^{2}}$, where the value of $\sigma_{\rm inst}$ ($\sim$1.34\,\AA) has been taken from the \textit{HST}/STIS handbook, and from \citet{Ho2009} for the Palomar data ($\sim$2.2\,\AA).


\section{Modeling of HST/STIS data}
\label{main_HST_results}

\begin{table*}  
	\caption{Results from the analysis of the optical spectra. From left to right: \lq ID\rq: object designation; \lq Inst.\rq: \textit{HST}/STIS or Palomar spectra; \lq Comp.\rq: components used for the best-fitting model with N, S and B referring to narrow, second and broad components, respectively; \lq [O\,I]\rq: if these lines are present in the spectrum or not (Yes/No); velocity (V) and velocity dispersion ($\sigma$) for each component for each of the reference emission lines; \lq f$_{\rm blend}$\rq: contribution of the broad component to the total flux of the H$\alpha$-[N\,II] complex.}
	\centering
	\begin{tabular}{l c c c c c c c c c c}
	\hline 						
ID & Inst. & Comp. & [O\,I] & V$_{\rm N}^{\rm [S\,II]}$ & $\sigma_{\rm N}^{\rm [S\,II]}$  &   V$_{\rm S}^{\rm[S\,II]}$ & $\sigma_{\rm S}^{\rm[S\,II]}$  &  V$_{\rm B}^{\rm H\alpha}$ & $\sigma_{\rm B}^{\rm H\alpha}$ & f$_{\rm blend}$ \\
   &   &    &       & (km\,s$^{-1}$) & (km\,s$^{-1}$) & (km\,s$^{-1}$) & (km\,s$^{-1}$) & (km\,s$^{-1}$) & (km\,s$^{-1}$) & (\%) \\
   (1)&(2)&(3)&(4)&(5)&(6)&(7)&(8)&(9)&(10)&(11)\\
 \hline  
NGC 2685 & \textit{HST}/STIS & N & N & 80\,$\pm$\,1 & 63\,$\pm$\,12 & - & - & - & - & - \\  
 & Palomar & N & Y & -10\,$\pm$\,1 & 70\,$\pm$\,10 & - & - & - & - & - \\ 
NGC 3245 & \textit{HST}/STIS & N\,+\,B & N & 134\,$\pm$\,29 & 172\,$\pm$\,6 & - & - & 424\,$\pm$\,12 & 998\,$\pm$\,6 & 47 \\ 
 & Palomar & N\,+\,S & Y & -216\,$\pm$\,9 & 25\,$\pm$\,40 & 94\,$\pm$\,17 & 136\,$\pm$\,21 & - & - & - \\
NGC 4374 & \textit{HST}/STIS & N\,+\,S\,+\,B & Y & -275\,$\pm$\,13 & 146\,$\pm$\,14 & 362\,$\pm$\,5 & 281\,$\pm$\,24 & 693\,$\pm$\,7 &  1432\,$\pm$\,37 & 65 \\
 & Palomar & N\,+\,S & Y & 117\,$\pm$\,4 & 51\,$\pm$\,29 & 33\,$\pm$\,56 & 395\,$\pm$\,71 & - & - & - \\
\textbf{NGC 4486} & \textit{HST}/STIS & N\,+\,S & Y & 395\,$\pm$\,4 & 198\,$\pm$\,22 & -101\,$\pm$\,378 & \textbf{655\,$\pm$\,18} & - & - & - \\
 & Palomar$^{*}$ & N\,+\,S\,+\,B & Y & -478\,$\pm$\,4 & 238\,$\pm$\,5 & 96\,$\pm$\,6 & 286\,$\pm$\,9 & -285\,$\pm$\,145 & 908\,$\pm$\,141 & 5 \\
NGC 4552$^{\dagger \ddagger}$ & \textit{HST}/STIS$^{\ddagger}$ & N\,+\,S\,+\,B & Y & -103\,$\pm$\,47 & 157\,$\pm$\,18 & 501\,$\pm$\,38 & 249\,$\pm$\,50 & 541\,$\pm$\,53 &  1360\,$\pm$\,20 & 70 \\
 & Palomar$^{\dagger}$ & N & Y & 56\,$\pm$\,59 & 282\,$\pm$\,68 & - & - & - & - & - \\
\textbf{NGC 4594} & \textit{HST}/STIS & N\,+\,S\,+\,B & N & 70\,$\pm$\,15 & 108\,$\pm$\,26 & 66\,$\pm$\,152 & \textbf{554\,$\pm$\,6} & 519\,$\pm$\,120 &  1677\,$\pm$\,18 & 16 \\
 & Palomar & N\,+\,B & Y & -11\,$\pm$\,4 & 215\,$\pm$\,7 & - & - & 698\,$\pm$\,26 & 956\,$\pm$\,41 & 33 \\
NGC 4676B$^{* \ddagger}$ & \textit{HST}/STIS$^{\ddagger}$ & N\,+\,B & N & 527\,$\pm$\,40 & 198\,$\pm$\,12 & - & - & 336\,$\pm$\,16 &  748\,$\pm$\,12 & 89 \\
NGC 4698 & \textit{HST}/STIS & N\,+\,B & N & 55\,$\pm$\,19 & 90\,$\pm$\,5 & - & - & 273\,$\pm$\,97 &  1375\,$\pm$\,9 & 39 \\
 & Palomar & N & Y & -7\,$\pm$\,2 & 98\,$\pm$\,8 & - & - & - & - & - \\
NGC 4736 & \textit{HST}/STIS & N\,+\,B & Y & 127\,$\pm$\,14 & 84\,$\pm$\,12 & - & - & 330\,$\pm$\,1 & 738\,$\pm$\,4 & 82 \\
 & Palomar & N & Y & 23\,$\pm$\,4 & 105\,$\pm$\,10 & - & - & - & - & - \\  \hline 
 	\label{T_kin}
	\end{tabular}
\tablefoot{The galaxies in bold show a secondary component in the \textit{HST}/STIS spectrum classified as outflow (see Sect.~\ref{disc_classification} and Fig.~\ref{Panel_vsig}). $^{*}$ indicates that the broad component detection in H$\alpha$-[N\,II] lines is not well constrained (Sect.~\ref{indiv_comments}). $^{\dagger}$ the data for which the O-model was used to model the emission lines (see Sect.~\ref{analysis_LF}). $^{\ddagger}$ marks the spectra with low S/N (Sect.~\ref{Sample_data}).} 
\end{table*}   
   
   \subsection{Overall modeling summary}
   Despite there were three possible models to fit the lines, we have selected the S-method to model all the \textit{HST}/STIS spectra for different reasons. For 5 LINERs the spectrum was lacking the [O\,I] lines either for the wavelength range coverage or because they were not detected (Table~\ref{T_kin} column 4). For two LINERs, NGC 4374 and NGC 4552, the S\,/\,N of the [O\,I] lines was rather faint and difficult to model (see Sect.~\ref{analysis_LF}). For NGC 4736, the fit presented no significant difference when applying S-/O-/M-models, so we decided to use the simplest model that could explain the observed spectra, in this case, the S-model. For the remaining object, NGC 4486, although one of the [O\,I] lines is clearly visible (Fig.~\ref{Panel_NGC4486}), the modeling was better with the S-method (the $\chi^{2}$ improved about $\sim$30\%). All the fits to the different spectra can be found in the Appendix~\ref{appendix}.
   
   Generally, the modeling of both [S\,II] and [O\,I] lines are inside the  3$\varepsilon_{\rm line}$ limits for all LINERs except for NGC 4374 (see Fig~\ref{Panel_NGC4374}). However, in the case of [N\,II]\,-\,H$\alpha$, $\varepsilon_{\rm c}$ is approximately equal to 3$\varepsilon_{\rm line}$ for NGC 4594 and $\sim$\,4$\varepsilon_{\rm line}$ for NGC 4374 despite the addition of the very broad component (Fig~\ref{Panel_NGC4594}). In all the other objects, $\varepsilon_{\rm c}$ is within 3$\varepsilon_{\rm line}$ limits, thus we consider the fit to be not well constrained for these 2 cases.

   We find that NGC 2685 is the only spectrum that can be modeled with a single narrow component in all the emission lines and NGC 4486 is the only spectrum well reproduced with a narrow plus a secondary component. No very broad component is required for them (Fig.~\ref{Panel_NGC2685} and~\ref{Panel_NGC4486}). 
   For the other 7 LINERs we found that 4 out of 9 need one narrow Gaussian component to model all the forbidden lines and an additional broad (BLR) component in H$\alpha$ (see Table~\ref{T_kin} column 3). 
   On the other hand, the remaining 3 LINERs need a secondary component to model the  forbidden lines and narrow H$\alpha$ in addition to a very broad component in H$\alpha$. 
   
   In summary, a secondary component is needed in 4 out of the 9 analyzed galaxies and a very broad component in H$\alpha$ is needed in 7 out of the 9 LINERs.

   \subsection{Comparison with Constantin et al. 2015 (C15)}
   \label{C15}

   \begin{figure}
   \includegraphics[width=\columnwidth]{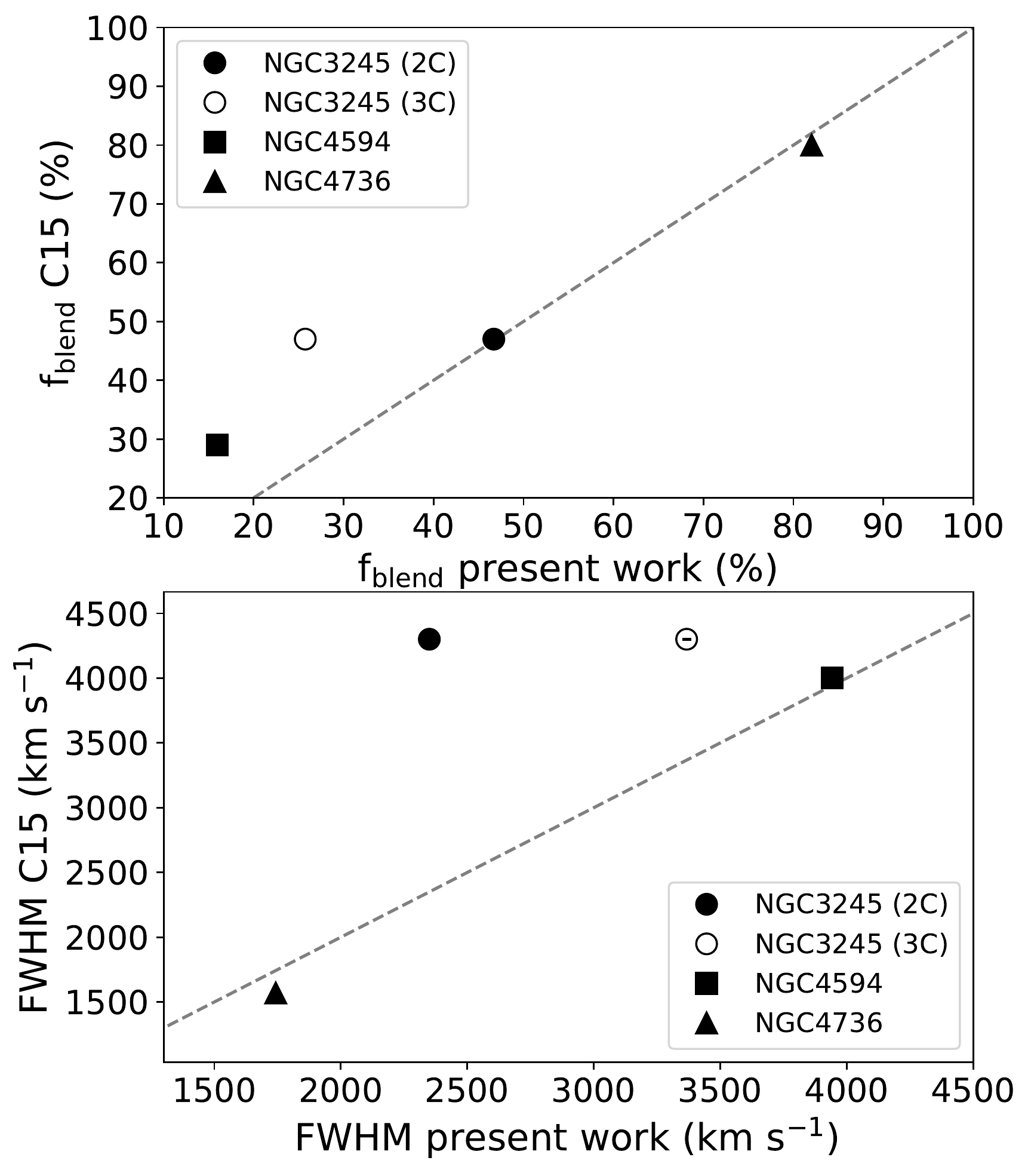}
   \caption{Comparison of the contribution in percentage of the broad component of H${\alpha}$ to the total flux of the H${\alpha}$-[N\,II] complex (upper panel) and the FWHM of the very broad component (lower panel) derived in this work versus C15. Each symbol represents a different LINER as indicated in the legend. The filled (empty) circle represents the two-(three-)component fit to NGC 3245. The gray-dashed line is the one-to-one comparison.}  
   \label{Panel_C15}
   \end{figure}

   Most of the galaxies shown here (7 out of 9; 77\% of the sample) have already been analyzed by C15, where they report for the first time the detection of broad components in 8 objects of their total sample. They have studied a total of 113 AGNs with \textit{HST}/STIS and MMT (Multiple Mirror Telescope) spectroscopic data to analyze the emission lines and to give constraints on the ionizing mechanisms present in the galaxies. The reduction process of the \textit{HST}/STIS data and the line modeling, for which they used the [S\,II] lines as template for the H$\alpha$-[N\,II] blend, follow a similar technique to us. 
   For the galaxies in common, they found a broad H$\alpha$ emission for three of them (NGC 3245, NGC 4594, NGC 4736) in agreement with our findings (see Table~\ref{T_kin} and  Fig.~\ref{Panel_NGC3245},~\ref{Panel_NGC4594} and~\ref{Panel_NGC4736}). 
   
   A direct comparison of the contributions of the very broad component to the H${\alpha}$-[N\,II] complex and the FWHM for these 3 galaxies can be seen in Fig.~\ref{Panel_C15}. NGC 3245 is represented with an empty or filled symbol depending on the number of components used for the modeling. Fig.~\ref{Panel_C15} shows that, although if we consider our two-component modeling the contribution of the very broad component to the H$\alpha$\,-\,[N\,II] complex is compatible in both analyses, the FWHM is more similar in the case of the three-component fit. The reason for this could be a direct consequence of the number of components. If we have two Gaussians for each emission line plus a broad component, its contribution to the global fit is going to be smaller than for a modeling of a single Gaussian plus a broad component, as the total flux of H$\alpha$\,-\,[N\,II] is going to be distributed among more components. 

   The number of components for the modeling of NGC 4594 and NGC 4736 is the same as C15 (see Table~\ref{T_kin}; Fig.~\ref{Panel_NGC4594},~\ref{Panel_NGC4736}), thus no more additional models were considered. In both cases the results are in agreement with C15, although these authors indicate that the very broad component of NGC 4594 is very difficult to interpret (see Sect.~\ref{ind_com_modelling_BLR}). 
   This very broad component dominates over the narrow to the H$\alpha$ line flux in both analyses, as can be seen in Fig.~\ref{Panel_NGC4594}. The broad component for NGC 4736 contributes significantly to the total flux of both H${\alpha}$ and the whole H${\alpha}$-[N\,II] complex, as can be seen in  Fig.~\ref{Panel_C15}.

   For three galaxies (NGC 4374, NGC 4552 and NGC 4698; Fig.~\ref{Panel_NGC4374},~\ref{Panel_NGC4552},~\ref{Panel_NGC4698}, respectively) we derive a very broad component ($>$\,1300\,km\,s$^{-1}$) that C15 do not report in their analysis. According to our modeling, the fit is improved in the three galaxies by adding a very broad H$\alpha$ component to the H$\alpha$-[N\,II] complex. These detections are individually discussed in Sect.~\ref{ind_com_modelling_BLR}.

   \section{Modeling of Palomar data}
   \label{main_palomar_results}
   
   \subsection{Overall modeling summary}
   The Palomar data was analyzed by \citet{Ho1997} (hereafter H97) also looking for the possible presence of a very broad component in the spectra. We have analyzed 8 galaxies of their sample by testing all the three different fitting methods as described in Section~\ref{analysis}. This is possible for the ground-based data-set, as the [O\,I] lines are visible in all cases (Table~\ref{T_kin}, column 4; see figures in Appendix~\ref{appendix}). 
   
   The best fitting is obtained with the S-method for nearly all the spectra (7 out of 8), as no significant improvement was seen when using the other methods. NGC 4552 was the only galaxy for which we selected the O-method due to the severe blending of the [S\,II] lines (see Fig.~\ref{Panel_NGC4552}).
   
   We find that 4 galaxies (NGC 2685, NGC 4552, NGC 4698 and NGC 4736) can be modeled with a single narrow component in all the emission lines and 3 galaxies (NGC 3245, NGC 4374 and NGC 4486) need a narrow plus a secondary component, with NGC 4486 requiring a very broad component to obtain a proper model of the spectrum (see Sect.~\ref{narrow_line_comments}). NGC 4594 is the only case whose spectrum is modeled with a narrow Gaussian component for all the forbidden lines and an additional broad H$\alpha$ component (see  Fig.~\ref{Panel_NGC4594}). 
   In summary, a very broad component in H$\alpha$ is needed in 2 out of the 8 galaxies and a secondary component is needed in 3 out of the 8 LINERs (see Table~\ref{T_kin}). 
   
   \subsection{Comparison with Ho et al. 1997 (H97)}
   \label{comparison_Ho}
   In the search for broad emission lines done by H97, they analyzed the data of the galaxies from \citet{Ho1995} whose line profiles showed (or at least hinted) the presence of a broad H$\alpha$ emission. This reduced the total sample from 486 to 211 galaxies, from which only in 34 objects this broad feature was detected. The possible existence of a broad component in eight of the common objects is discussed below.

   The methodology used in H97 to analyze the spectra consisted on an initial stellar subtraction using the spectra of galaxies devoid of emission lines (i.e. \lq template-galaxies method\rq, \citealt{Ho1993} and references therein) and then a line modeling.  C18 also tested the \lq template-galaxies method\rq, although when applying the \textsc{ppxf} routine to the spectra of the templates (see Sect.~\ref{analysis}), some weak emission lines were visible in 6 out of the 10 template galaxies. 
   Thus, we used the \textsc{ppxf} method instead of the \lq template-galaxies method\rq , as they need to be treated very carefully in order to avoid the inclusion of emission lines produced by the interstellar medium (see C18 for further details).
   
   The line modeling in H97 consisted on a single Gaussian fit to the [S\,II] lines, considering them as reference to model the H$\alpha$-[N\,II] lines. Similar to our methodology, an additional broad component could be added to the H$\alpha$ line only when necessary (see Sect.~\ref{analysis_LF}). 
   
   The spectra modeling for 4 of the galaxies (NGC 3245, NGC 4486, NGC 4594 and NGC 4698) is shown in H97. Three of them (namely NGC 4552, NGC 4594 and NGC 4698) are among their null detection of a broad component, whereas NGC 2685, NGC 3245 and NGC 4486 are among the ambiguous cases. For these, according to H97, the cause of ambiguity is the template chosen for the starlight subtraction.
   
   In our analysis, thanks to the use of \textsc{ppxf}, we minimized possible source of ambiguity related to the stellar modeling and subtraction (see Sect.~\ref{analysis} and C18). In our modeling for these same spectra, we find a broad component in 2 out of the 8 common objects (NGC 4486 and NGC 4594). Only for NGC 4594 this broad component is also visible on the \textit{HST}/STIS spectra (see Fig.~\ref{Panel_NGC4594}). This result indicates that the two objects, or at least NGC 4594, could have been initially classified as type-1 LINERs also with ground-based spectroscopy using a different starlight decontamination method and a different line modeling. 
   For the other galaxy, NGC 4486, the existence of a broad component has been a matter of debate in previous works with both ground- and space-based spectroscopy \citep[e.g.][ see Sect.~\ref{narrow_line_comments}]{Harms1994, Ho1997, Walsh2013, Constantin2015}. It was described in H97 as showing weak wings at either side of [N\,II] lines. Although when adding a broad component to the fitting their residuals improve, they say that this  component could be a product of some ambiguities with the modeling, so they decided not to include it in the analysis. Its contribution to the global fit is rather weak (5\%), but we also improve the residuals from our fitting by including it (Fig.B4) . We discuss its presence or absence in our ground and space data in Sect.~\ref{BLR}.


\section{Results and Discussion}
\label{discussion}

   We note that, given the difference among the slit-widths between Palomar and \textit{HST}/STIS data (see Sect.~\ref{analysis_LF}), we will consider the latter results as the most accurate in characterizing the innermost parts of the AGN. Hence, the following kinematical analysis is only focused on them. 
   Apart from this, the possible discrepancies between the measured narrow-component velocities from ground- to space-based data (see Table~\ref{T_kin}) may be produced by the different alignment of the slits.

   \subsection{Overall summary of kinematics}
   \label{summary_kin_lineratios}

   The wavelength-shift and line-width of each of the Gaussian components used to model the spectra are converted into a measurement of velocity and velocity dispersion of the ionized gas kinematics. We can associate them to a kinematic component of the galaxy. To this aim, we have followed the velocity-velocity dispersion (V-$\sigma$) classification similar to that used in C18, which is shown in Fig.~\ref{Panel_vsig}. 
   
   This figure is divided into 4 different regions, associated to rotation, outflows, inflows and candidates for rotational motions. Dividing limits of the regions have been established by C18 by measuring the gas velocity field from the 2D spectra of the type-1.9 LINERs within their sample, and estimating the maximum broadening of the emission lines associated to rotational motions. 
   The C18 conservative upper limit for this component is $\sim$400\,km\,s$^{-1}$, although the typical values of velocity dispersion ($\sim$200\,km\,s$^{-1}$) in their sample is well below this limits (see Sect.~5.2 in C18). This limits are supported by the fact that the ionized gas in rotation usually produces a broadening of the lines that is traduced into velocity amplitudes up to  $\sim$300\,km\,s$^{-1}$ \citep[see e.g.][] {Cappellari2007,Epinat2010}. They found that the regions outside the nuclear zone in the 2D spectra had amplitudes $\sim$300\,-\,400\,km\,s$^{-1}$, thus associating this part of the V-$\sigma$ diagram to candidates for broadening produced by rotational motions.
   
   As for the velocities, typically the narrow component of type-1.9 LINERs has values $\pm$50\,km\,s$^{-1}$. Thus, when the $\sigma$ is larger than 300\,km\,s$^{-1}$, these broadening cannot be addressed to rotational motions but instead to non-rotational motions. Then we would consider them as candidates to the presence of outflows or inflows. For the intermediate cases, the targets are considered as candidates for non rotational motions. 
   
   For the secondary component, C18 found that the velocity dispersion present typically values from 400 to 800\,km\,s$^{-1}$. In this case, the broadening is probably related to turbulent non-rotational motions. It is seen in Fig.~\ref{Panel_vsig} that the velocities for the secondary component of type-1.9 LINERs present typical values ranging from -200 $\leq$ V$_{S}\leq$ 100\,km\,s$^{-1}$ (see C18).
   For this analysis we have decided to apply similar limits to C18 because the complete study of the gas velocity field could not be done with the present data-set. Thus, it has to be noted that the region limits could be different for type-2 LINERs if a similar analysis is considered. However, the physical interpretation of the position in the diagram for the galaxies with a secondary component is unlikely to change dramatically given their large velocity dispersion $>$500\,km\,s$^{-1}$ and hence their position on Fig.~\ref{Panel_vsig}.
   
   \subsection{Classification of the velocity components}
   \label{disc_classification}

   \begin{figure*}
   \includegraphics[clip,trim=0.5cm 11.8cm 1.4cm 9cm,width=\textwidth]{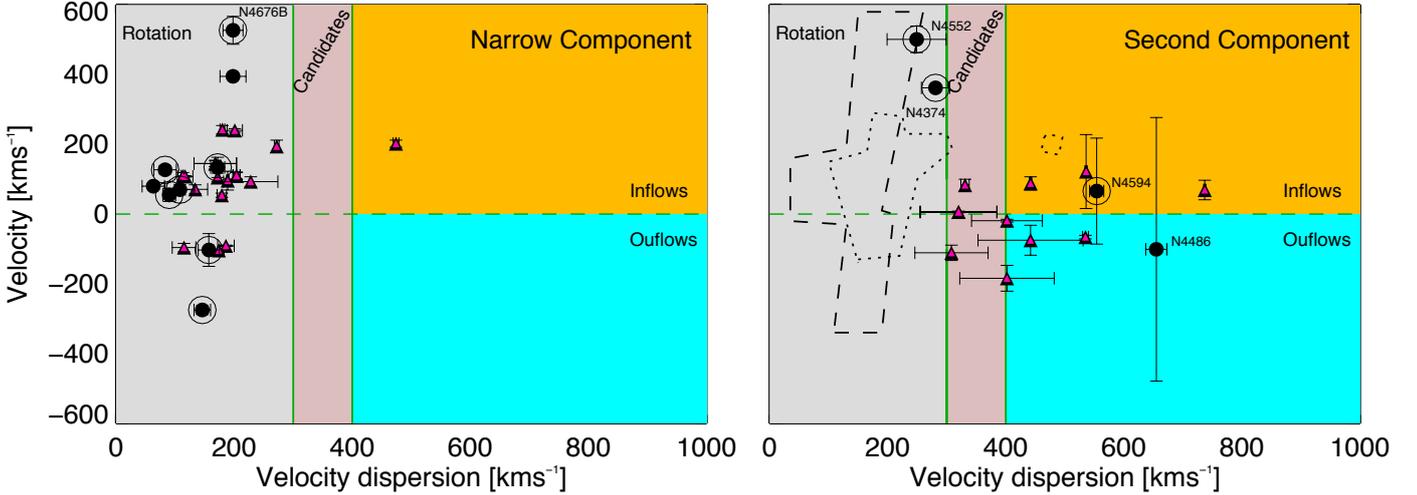}
   \caption{Velocity dispersion versus velocity derived for the narrow (left panel) and secondary (right panel) components used to model the \textit{HST}/STIS spectra for both type-1.9 from C18 (red triangles) and type-2 (black filled circles) LINERs. An additional black circle marks those cases for which a broad component is needed in \textit{HST}/STIS spectra. All type-1.9 LINERs from C18 show a broad component. 
   Each of the marked regions indicate to which kinematic component can be associated the modeled component. The orange region indicates inflows, blue region indicates outflows, pink region indicates candidates for both inflows and outflows, and the gray region indicates rotation. On the right panel, the contours of the narrow components are shown for both type-1.9 (dotted lines) and type-2 (dashed lines) LINERs.}
   \label{Panel_vsig}
   \end{figure*}

   As mentioned in the previous section, if the velocity dispersion is in the range up to 300\,km\,s$^{-1}$, the component may be associated to the presence of a rotating disc in the galaxy. The broad H$\alpha$ component is not considered in this figure as it is interpreted unequivocally as a manifestation of the BLR of the AGN. The narrow component, present in all the ground- and space-based spectra, shows velocity dispersions consistent with being produced by rotation. On the other hand, the secondary component is always broader than the narrow, and has a higher spread on the velocity dispersion for both type-1.9 and type-2 LINERs, with typical values higher than $\sim$\,400\,km\,s$^{-1}$. These values are consistent with the presence of non-rotational motions (outflows or inflows). 

   In Fig.~\ref{Panel_vsig}, only four galaxies of the \textit{HST}/STIS sample (NGC 4374, NGC 4486, NGC 4552 and NGC 4594) needed a secondary component on the line modeling. From these, only for NGC 4374 and NGC 4486 a second component is added in the Palomar spectra. NGC 4486 and NGC 4594 line widths of the secondary component are consistent with being caused by the presence of non-rotational motions, which is also suggested by previous results \citep[e.g. the H$\alpha$ morphology of NGC 4486 shown in][]{Masegosa2011}. 
   For the remaining cases, the line widths are compatible with being produced by rotational motions, although NGC 4374 could be a possible inflow candidate. Also in \citet{Masegosa2011} the possible presence of an outflow seen with the H$\alpha$ emission was discarded (or inconclusive) for this galaxy due to the dust lanes that obscure and limit the access to the nuclear H$\alpha$ morphology. 
   For NGC 4552, a bipolar nuclear outflow was reported by \citet{Machacek2006} with \textit{Chandra} images (0.5\,-\,2 keV). 
   
   The other galaxy in the sample which has been previously reported as hosting an outflow is NGC 3245. The H$\alpha$ image of this galaxy shows an unresolved nucleus and a structure indicative of the presence of an outflow \citep{Masegosa2011}, which is also supported by the kinematic analysis of \citep{Walsh2008}. However, in our modeling of the \textit{HST}/STIS spectrum, a secondary component is not needed (see Sect.~\ref{narrow_line_comments}) and, if used, the resulting velocity and velocity dispersion will be compatible with being caused by rotation (V$_{\rm s} = -40\pm25$\,km\,s$^{-1}$; $\sigma_{\rm s} = 194\pm56$\,km\,s$^{-1}$). 
   With the Palomar data instead, a secondary component is clearly required to satisfactorily reproduce the line profiles with low residuals (Fig.~\ref{Panel_NGC3245}). In this case, the secondary component also has a velocity and velocity dispersion in agreement with rotation, so it is not clear that it could be associated to non-rotational motions. 

   We can directly compare our values with the results from the \textit{HST}/STIS analysis of 12 type-1.9 LINERs from C18. It is clearly seen that the narrow component for both samples can be explained by rotational motions (see Fig.~\ref{Panel_vsig}). The typical values of velocities and velocity dispersion (see Sect.~\ref{summary_kin_lineratios}) are similar except for some cases. NGC 4676B has a velocity higher than 500\,km\,s$^{-1}$. It is a merging galaxy, and its modeling is not well constrained (see Fig.~\ref{Panel_NGC4676B} and Sect.~\ref{indiv_comments}). In \citet{Wild2014} they quote a big velocity gradient oriented in a similar direction to where the \textit{HST}/STIS slit is allocated, which may suggest that the velocity of the narrow component could be affected by this. NGC 4374 and NGC 4486 have larger velocities of the narrow component than what should be expected (see Sect.~\ref{summary_kin_lineratios}). For the latter also \citet{Noel2003} found that the narrow component was not rest frame in the central spectrum. For the first, they might be produced by the large velocity gradients present in the innermost parts and that we are not considering individually in this analysis (see Sect.~\ref{ind_com_modelling_BLR}).
   On the other hand, the values for the secondary component are more spread throughout the diagram, including 2 galaxies in the range in which the broadening could be explained by rotation. Velocities for the secondary component of type-1.9 LINERs are more concentrated than type-2 LINERs, with values ranging from -110\,$\leq$\,V$_{S}$\,$\leq$\,500\,km\,s$^{-1}$ for the latter. The velocity dispersion range is more similar for both samples, ranging from 250\,$\leq$\,V$_{S}$\,$\leq$\,800\,km\,s$^{-1}$. The proportion of objects within the C18 sample that present a secondary component is slightly higher than that of type-2 LINERs (7 out of 11 versus 4 out of 9). This result could be a consequence of the orientation of the slit, that could not capture the full extension of putative outflows or inflows. However, this comparison could be not completely representative of these populations due to the modest amount of type-2 LINERs in the sample. 

\subsection{Individual comments for the galaxies}
   \label{indiv_comments}

   This section is organized taking into account the components used to fit the \textit{HST}/STIS spectra, as indicated in Table~\ref{T_kin} (column 3). The figures with the spectra modeling are in Appendix~\ref{appendix}.
   As in all cases the BLR components in our analysis have large velocity shifts, we have tried to set the velocity of the BLR-component to zero in order to statistically confirm its true velocity. The majority of the models presented worse residuals both visually and statistically (i.e. difference in $\chi^{2} > 15$\% worse) than with the final velocities (see Table~\ref{T_kin}). We discuss individually the results for the galaxies that did not change much with this model.

    \subsubsection{Model with only narrow components}
    \label{narrow_line_comments}

    As mentioned in Sect.~\ref{main_HST_results}, we have been able to model only two of the galaxies (NGC 2685 and NGC 4486) with a narrow or a narrow plus a secondary component (see Table~\ref{T_kin}). 
    
    The \textit{HST}/STIS spectra of \textbf{NGC 2685} has been analyzed in two works. \citet{GD2004} compare the main ionizing mechanism of this LINER measured with both ground- and space-based instruments, although the data-set for STIS use the G430L grating. It is the only galaxy in their analysis that requires a single narrow Gaussian component to reproduce the observed spectrum. C15 applied exactly the same modeling for the same data-set using only narrow components. Hence, our results are consistent with previous ones.

    For \textbf{NGC 4486} is a complex galaxy with a well-known radio jet. \citet{Noel2003} analyzed this spectrum in first place and found that the presence of a broad component improved visually and statistically the global fit of the spectrum with respect to a single Gaussian component per line. The velocity of the narrow component they measure in the central pixel is $\sim$ 260\,km\,s$^{-1}$. Neither \citet{Walsh2013} nor C15 found a broad H$\alpha$ component when modeling the emission lines. \citet{Walsh2013} reports the existence of velocity gradients in the innermost regions of the slit, as for the case of NGC 4374 (see Sect.~\ref{ind_com_modelling_BLR}). The velocity dispersion can be as high as 600\,km\,s$^{-1}$ in the central row, and it decays in the following regions. In our case, this structure is integrated in our spectrum.
    Our fit does not improve significantly by adding a very broad component to a two component fitting (only a 5\%), thus only two components were used. In this case, the secondary component is consistent with an outflow (see Table~\ref{T_kin} and Fig.~\ref{Panel_vsig}). However, a second component with large width could be also the result of the strong  gradient in the velocity dispersion reported by \citet{Walsh2013}. For the Palomar spectrum, a broad component improves a 18\% the fit, although the $\varepsilon_{\rm c}$ was already smaller than 3$\varepsilon_{\rm line}$ (see the residuals of the fit without the broad component in Fig.~\ref{Panel_NGC4486}). Given this and the non-clear detection in previous works, we should consider the (non) detection in the Palomar (\textit{HST}/STIS) spectrum as ambiguous. The non-detection of the \textit{HST}/STIS data implies that this component, if existing, should not be attributed to typical BLR clouds orbiting nearby the black hole but instead to other phenomena on larger scales (e.g. inflows/outflows).

    \subsubsection{Models with broad BLR-originated component}
    \label{ind_com_modelling_BLR}
    
    The remaining 7 galaxies were found to require a very broad component with the emission-line fitting analysis. 
    
    \textbf{NGC 3245} was classified as an intermediate case between a LINER and a transition object \citep{Ho1997}. \textit{HST}/STIS data was analyzed in first place by \citet{Barth2001}, who studied the dynamics of the ionized gas in this galaxy. 
    They suggested the possible presence of a broad component in H$\alpha$, which was modeled by C15 (see Sect.~\ref{C15}). Their emission line modeling consisted on 3 Gaussian profiles fitted to [N\,II]\,-\,H$\alpha$ lines (tied in velocity but allowed to vary in $\sigma$). 
    In our case, visual inspection indicates that only one narrow component is sufficient to fit the lines, although with a secondary component the $\chi^{2}$ of the fit improves approximately a 16\%. We have prioritized the simplest model that is able to describe the observed spectrum (S-method; see Fig.~\ref{Panel_NGC3245}). In either case, a broad component in H$\alpha$ was necessary to model the line profiles. 

    The spectrum of \textbf{NGC 4594} obtained with the TIGER 3D spectrograph shows some irregularities in its nucleus and a strong velocity gradient ($\sim$300\,km\,s$^{-1}$) near the center \citep{Emsellem2000}. 
    However, as neither the HST or Palomar slits are located along the major kinematic axis of the galaxy (see Fig.~\ref{Panel_NGC4594}), we would not expect a strong line-broadening due to the ordinary disk rotation. The H$\alpha$ emission is concentrated in the spiral arms and in the nuclear region of the galaxy \citep{Masegosa2011}. 
    There has been some debate about the presence of a broad H$\alpha$ component in the spectra of this galaxy. There are studies supporting both its presence \citep{Kormendy1996,Walsh2008} and its absence \citep{Ho1997,Nicholson1998}. Hence a methodology that takes into account all the different possibilities, as the one presented in this work, is required to understand the true nature of this LINER-nucleus. The spectrum presents wings in the H$\alpha$\,-\,[N\,II] lines that was initially attributed to the existence of a BLR with data from the \textit{HST}/Faint Object spectrograph \citep{Nicholson1998}. On the one-to-one row analysis by \citet{Walsh2008}, they show that there are large velocity gradients in the nuclear regions covered by the \textit{HST}/STIS slit. These gradients could be producing the observed line-profile, as indicated in Sect.~\ref{analysis_LF}.
    The best fit to the \textit{HST}/STIS spectrum is obtained with two components plus a broad component, as C15 (see Sect.~\ref{C15} and Fig.~\ref{Panel_NGC4594}). The residuals are still higher than 3$\varepsilon_{line}$, they have been reduced from 3.2 to 2.7 for [N\,II]$\lambda$6548\AA\, and the $\chi^{2}$ of the overall fit is improved a 15\% by adding the broad component. 
    If we were to set the velocity of the BLR component to zero, the $\chi^{2}$ would be only a 3\% worse and the $\sigma$ would be 10\,km\,s$^{-1}$ less broad. However, as this component has a negligible flux contribution to the total H$\alpha$-[N\,II] blend and it is the broadest line we found in our study (see Table~\ref{T_kin}), it should be expected to find no difference between both modelings statistically. However, visually the red wing of the [N\,II]6584$\AA$ line is better reproduced with the velocity as indicated in Table~\ref{T_kin}. Hence, we decided to not set to zero the velocity of the broad H$\alpha$ component.
    
    \textbf{NGC 4736} is an interesting case as the broad component is clearly visible in the \textit{HST}/STIS spectrum \citep[][C15]{Constantin2012} but it has never been reported from ground-based observations \citep{Ho1997,Constantin2012}. The  size of the slit-aperture is crucial in this object to classify it properly. 
    The broad component has also been reported with a Principal Component Analysis (PCA) Tomography method applied to Integral Field Unit (IFU) data taken several years after the \textit{HST}/STIS spectra \citep{Steiner2009}. 
    In our analysis all the forbidden lines, except the H$\alpha$-[N\,II] complex, are  well-modeled with a single narrow Gaussian component. For H$\alpha$ a broad component of $\sigma_{\rm B} \sim 740$\,km\,s$^{-1}$ is needed (see Sect.~\ref{C15} and Fig.~\ref{Panel_NGC4736}). The FWHM of this component is in the limit of being associated with the BLR of the galaxy, which is also the case for NGC 4676B (see below). Both $\sigma_{\rm B}$ are only slightly higher ($\sim$100\,km\,s$^{-1}$) that the $\sigma_{\rm S}$ for NGC 4486.
    
    C15 did not report additional very broad detections for the other galaxies in common (NGC 2685, NGC 4374, NGC 4486 and NGC 4552; see Sect.~\ref{C15}). However in our analysis two of them present a very broad component, whereas the other two only need a narrow and a narrow plus a secondary component (NGC 2685 and NGC 4486, respectively), as indicated previously (see Sect.~\ref{narrow_line_comments}). 
    
    For the modeling of NGC 4374 and NGC 4552 nuclear spectra we have used a different technique from the rest of the galaxies. As explained in Sect.~\ref{analysis_LF}, in both cases the [O\,I] lines are barely detected, hence only the S-method could be applied but [S\,II] are severely blended.
    
    The \textit{HST}/STIS spectra of \textbf{NGC 4374} were first analyzed by \citet{Bower1998} and a few years later by \citet{Walsh2010}. In the former work, they used two Gaussian components to fit the lines, considering that they were tracing two dynamically different gas components, (a rotating disk and a low-velocity component unrelated to the disk). The latter showed that the profiles of [N\,II]-H$\alpha$ could be affected by the NLR kinematics, which is rapidly rotating, although their analysis do not exclude the presence of a broad component. 
    As in the case of NGC 4594, the presence of a BLR-originated component is debated. In C15 they concluded that a broad component in H$\alpha$ was not needed (see Sect.~\ref{C15}). Nevertheless, its presence was not ruled out by \citet{Walsh2010} in the 3 innermost rows, where a peaked central velocity dispersion, associated to the emission-line disk, is related to large velocity gradients visible mainly in the central row of the slit (see Sect.~\ref{disc_classification}). In our analysis, these gradients are integrated in the final spectrum as we do not perform a one-to-one row analysis (to maximize the S/N). In this sense, our narrow component fit should be properly considered as a lower limit to the velocity dispersion of the disk. Both narrow and secondary component are compatible with rotation (see Fig.~\ref{Panel_vsig} and Table~\ref{T_kin}), which is somewhat expected from the results of \citet{Walsh2010}. In this work, they do not show the values of the three nuclear rows analysis, as their kinematical profiles are said to be complicated to analyze due to the large gradients. However the amplitude of the velocities is comparable with that of the secondary component ($\sim$400\,km\,s$^{-1}$), being coincident with assumed limit for detecting outflows (Sect.~\ref{disc_classification} and Fig.~\ref{Panel_vsig}).
    
    In the case of \textbf{NGC 4552}, \citet{Cappellari1999} modeled the emission lines with intermediate width profiles in both permitted and forbidden lines and highlighted the low luminosity of the central engine in comparison with others AGNs. With respect to the \textit{HST}/STIS spectra, C15 do not consider a broad component. 
    In our modeling, that take into account a second component in forbidden lines and narrow H$\alpha$, reveals the presence of a broad component in both galaxies that contributes significantly ($\sim$ 65\% and $\sim$ 70\%, respectively) to the global fit, as can be seen in Figures~\ref{Panel_NGC4374} and~\ref{Panel_NGC4552} (see Sect.~\ref{C15}). 
    
    \textbf{NGC 4698} data come from the Survey of Nearby Nuclei with STIS, known as SUNNS, analyzed by \citet{Shields2007}. They suggest that the galaxy was badly classified as a type-2 Seyfert galaxy by \citet{Ho1997}, on the basis of a decrease of the level of the excitation mechanism, suggesting that it is a LINER or a transition object. These authors did not detect a broad component in H$\alpha$ in their spectral analysis, as \citet{Balmaverde2013} and C15 when analyzing the same data-set in contrast to our finding. 
    More specifically, we have found that, even though the [S\,II] lines are perfectly modeled with a single narrow component, the addition of a weak broad component improves the $\chi^{2}$ of the fit a $\sim$\,17\% (see Fig.~\ref{Panel_NGC4698}). The lack of detection of this component in previous works could be due to its low contribution to the global fit (39\%, Table~\ref{T_kin}) flux in comparison with the narrow component (see Sect.~\ref{C15}). 
    If we were to set the velocity of the broad component to zero, we find that the $\chi^{2}$ of the fit would be only 0.5\% worse than the fit with the velocity from Table~\ref{T_kin}. As for NGC 4594, the broad component has a rather low contribution to the global flux and is the second broadest component in our analysis, thus again statistically we should expect a small difference. In this case, also visually there was no difference. However, due to the rather low velocity of this component in the previous model (273$\pm$97\,km\,s$^{-1}$; Table~\ref{T_kin}), we decided not to add any constrain on the broad component velocity.
    
    The remaining galaxy, \textbf{NGC 4676B}, belongs to the Mice galaxies complex, two interacting spiral galaxies. Both spirals have been told to present features indicative of the presence of outflows \citep[][]{Masegosa2011, Wild2014}. Its \textit{HST}/STIS nuclear spectrum has not been studied before as the S/N is rather low, which makes it difficult to analyze (see Fig.~\ref{Panel_NGC4676B}). Despite this, we attempted to model the emission line, being the unique measurement of this kind for this object with \textit{HST}/STIS data.
    Our fit has revealed the presence of a broad component for the H$\alpha$-N[\,II] complex, while a narrow component is sufficient to fit the [S\,II] lines. Even though the $\chi^{2}$ improved with the addition of this component, the residuals are inside the 3$\varepsilon_{\rm line}$ limits also with a single narrow-modeling for all the lines. The presence of this component could be affected by the noise and the FWHM of this component is not as broad (748\,km\,s$^{-1}$; Table~\ref{T_kin}) as the other detections in our sample. The results of the test of setting the velocity of the broad component to a zero velocity show the same results in FWHM and $\varepsilon$-indicator. Then it is not clear if it can be associated to the presence of a BLR. To avoid possible ambiguities, in what follows we have conservatively excluded it as a true detection of a broad component.

  \subsection{BLR in optically classified type-2 LINERs?}
  \label{BLR}
  
  Type-2 LINERs have been classified by optical, ground-based spectroscopy as not having a broad component in the Balmer emission lines while it is not the case for type 1-1.9 \citep{Ho1993,Ho1995}. Nevertheless we have seen in our analysis (Sect.~\ref{indiv_comments}) that 6 out of the 9 analyzed galaxies in our sample exhibit this component in space-based spectroscopy. In the case of the 22 type-1.9 LINERs (selected from the Palomar Sky Survey) all the available space-based spectrum (12 LINERS) showed a broad component in H$\alpha$ (C18 for details).
  
  However, the same objects analyzed with the original Palomar spectra from \citet{Ho1997} do not show this feature except for 2 cases in type-2 LINERs (NGC 4486 and NGC 4594), and do show it in all except 4 cases in type-1.9 LINERs. 
  Ho's initial classification is mainly supported by this result. However, this enhances the difference of the ground-based spectra with those from space, based also on the results of C18. 
   
  The detection of a broad component is favored in space-based spectra because of the spatial resolution \citep{Molina2018}. If the BLR exists in a particular galaxy, a wider slit on a ground-based instrument could prevent it to be detected, as the weak broad H$\alpha$ component could be diluted. This could be the reason for the initial lack of detection of the BLR in the galaxies of our sample, but also the modeling of the lines (e.g. testing single and multiple components) could be crucial to highlight its presence (as it was our case for \textit{HST}/STIS spectra of NGC 4374 and NGC 4552).
  
  NGC 4486 is the only case in which the broad component is detected on the Palomar spectrum but not in the \textit{HST} data. As we should expect exactly the opposite behavior (i.e. the detection of this component on \textit{HST} and not in Palomar spectrum), it is not clear if this galaxy should be reclassified (type-1 instead of type-2). Moreover if we take into account that the existence of the BLR in this object has been largely discussed in the literature (see Sect.~\ref{comparison_Ho}). 
  On the other hand, NGC 4594 is the only object in which a very broad component is detected in both data-sets, so it should then be considered as a type-1 LINER. 

  \subsection{Line ratios} 
  \label{line_ratios}
  
  \begin{figure}
     \includegraphics[width=\columnwidth]{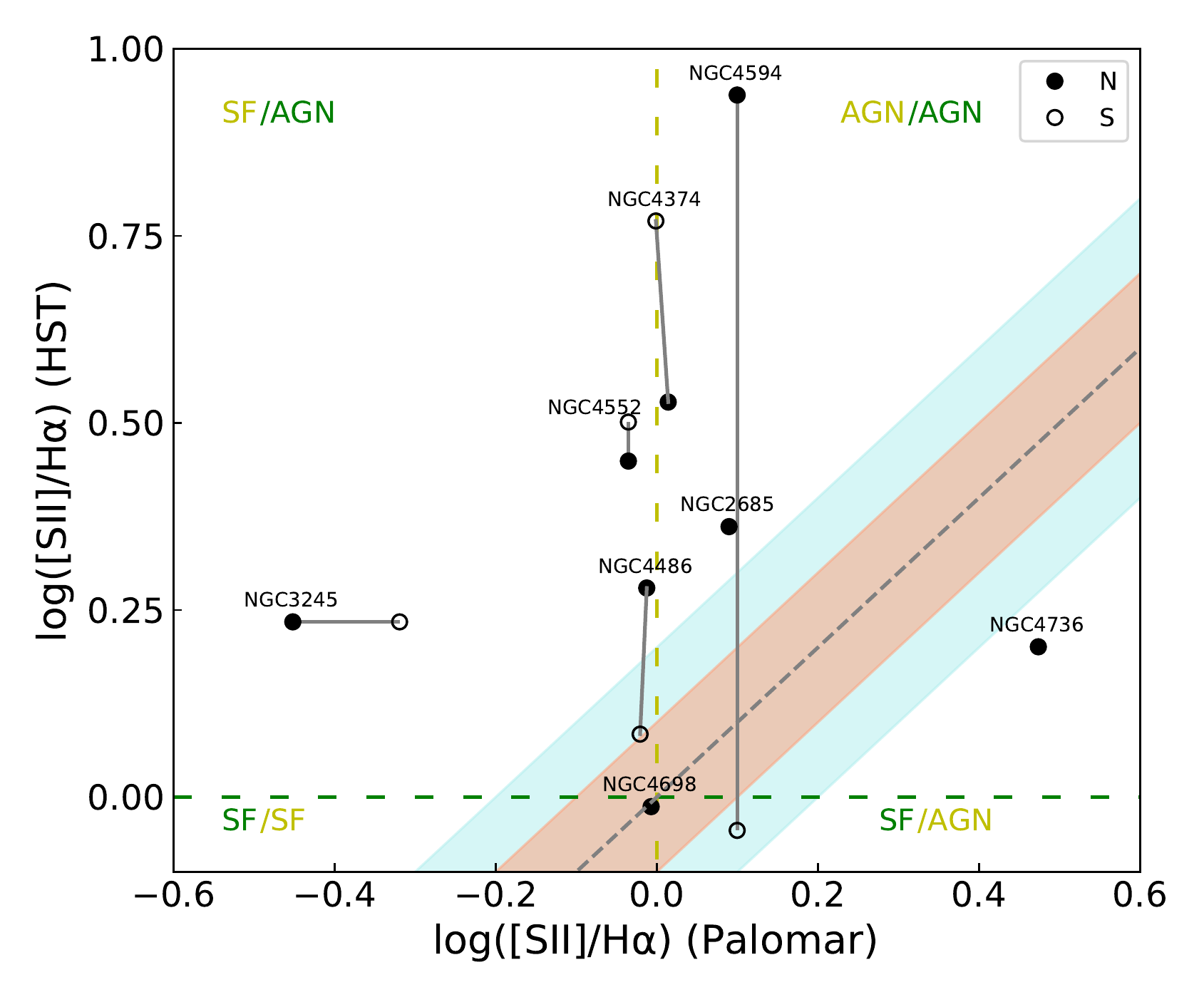} 
     \includegraphics[width=\columnwidth]{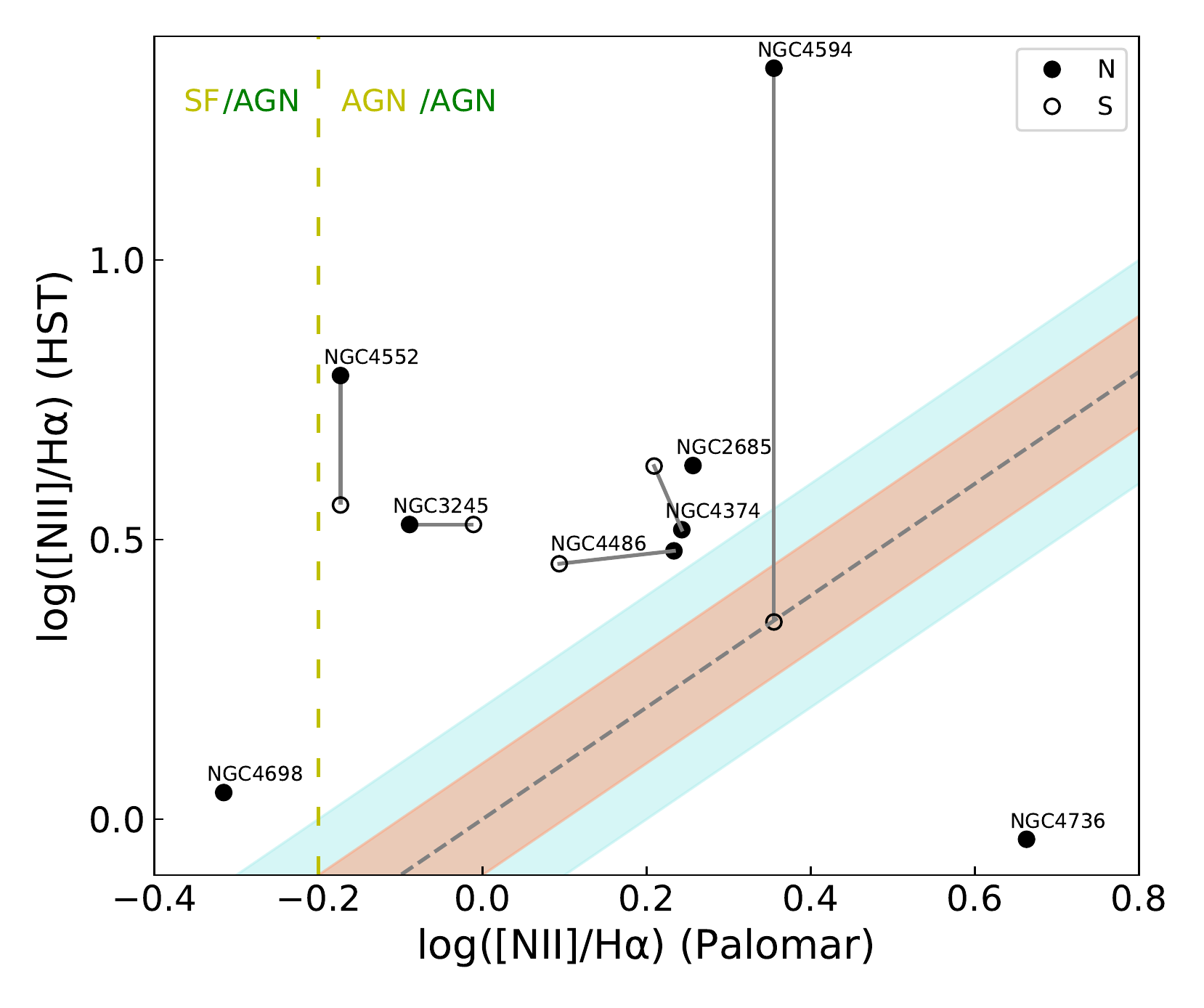} 
      \caption{Comparison of the line ratios for [S\,II]/H$\alpha$ (top) and [N\,II]/H$\alpha$ (bottom) for both \textit{HST}/STIS spectra (y-axis) and Palomar spectra (x-axis). The gray dashed line is the one-to-one comparison. Filled circles represent the narrow component whereas the empty circles are for the secondary, which is connected to the narrow through a gray solid line when present. The light pink and light blue ranges indicate the 10\% and 20\% uncertainty levels with respect to the one-to-one comparison, respectively. Green and yellow dashed lines indicate the limits (\textit{HST} and Palomar, respectively) for the star-forming (SF) and AGN regions. The limits were obtained from \citet{Kauffmann2003} and \citet{Kewley2006}.}
     \label{BPTs} 		 		 
  \end{figure}

  We have analyzed the line ratios for the emission lines available in all spectra. 
  In Figure~\ref{BPTs} the [N\,II]/H$\alpha$ and [S\,II]/H$\alpha$ line ratios for both Palomar and \textit{HST} data are shown\footnote{[O\,I]/H$\alpha$ is not considered as the [O\,I] line is not available for all \textit{HST} spectra (Table~\ref{T_kin}, column 4)}. We have estimated the ratios for both the narrow and the secondary components (when present) for all galaxies except NGC 4676B, as its modeling is not well constrained (see Sect.~\ref{indiv_comments}). 

  We consider in Fig.~\ref{BPTs} the empirically-derived separation between an AGN or a star forming (SF) system as in the BPT diagrams \citep[][]{Baldwin1981,Kauffmann2003,Kewley2006}. 
  Based only in [N\,II]/H$\alpha$ and [S\,II]/H$\alpha$ ratios. Since [O\,III]/H$\beta$ is not available in our data, we consider the star formation as the dominant mechanism of ionization if log([S\,II]/H$\alpha$) is $<0.0$. The mechanism would be the AGN ionization when log([S\,II]/H$\alpha$)$>0.0$. For the log([N\,II]/H$\alpha$), the limit for SF is $<-0.2$, for AGN $>0.2$, and  \lq transition\rq \ region elsewhere.
  
  In general, evident differences are seen in both panels of Fig.~\ref{BPTs} between the space- and ground-based data. For both [N\,II]/H$\alpha$ and [S\,II]/H$\alpha$, the ratios are always positive for the HST data, whereas more dispersion in the values exists in Palomar data (Palomar vs \textit{HST} mean/standard deviation for [N\,II]/H$\alpha$ 0.15/0.30 vs 0.54/0.40; and [S\,II]/H$\alpha$ 0.02/0.24 vs 0.37/0.26). Specifically, for Palomar (\textit{HST}) the average and standard deviation are for log([N\,II]/H$\alpha$) equal to 0.15 and 0.30 (0.54 and 0.40), respectively; for log([S\,II]/H$\alpha$) these values are 0.02 and 0.24 (0.37 and 0.26).
 
  In absence of [O\,III]/H$\beta$ (see Sect.~\ref{Sample_data}), the ratios are more compatible with the AGN region for the \textit{HST} spectra than for the Palomar data. The difference could be due to the spatial resolution based on the recent results from \citet{Molina2018}. They have analyzed 3 LINERs with \textit{HST} data, and performed a detailed study about the line diagnostic diagrams considering the unresolved nuclear source and what they call an \lq integrated\rq\, spectrum, that resembles the spectra of the targets as seen from a ground-based telescope. They found that the line ratios tend to be higher for the unresolved nuclear source than for the integrated spectrum. They conclude that it could be produced by the addition of a large spatial region of the galaxy ($\sim$100 pc) within the integrated spectrum, where there is influence from gas probably suffering simultaneously photoionization and shock effects. This is reflected on their line ratios being consistent with different ionization mechanisms.  

  In Fig.~\ref{BPTs}, NGC 2685, NGC 4698 and NGC 4736 only have a single narrow component (black circles), so they appear only once. The difference with respect to the one-to-one line could be explained directly with the variation of the H$\alpha$ profile seen in the \textit{HST} with respect to the Palomar spectra (see Fig.~\ref{Panel_NGC2685}, \ref{Panel_NGC4698} and~\ref{Panel_NGC4736}). Although, except for NGC 4698, they are compatible with being produced by an AGN in both Palomar and \textit{HST}. For the case of NGC 4698 line ratios correspond to the transition region in both diagrams for the \textit{HST} data, so this results in an ambiguous classification. The varying H$\alpha$ profile does not only produce considerable differences in the line ratios for these 3 objects, but also for NGC 3245 and NGC 4486 (see Fig.~\ref{Panel_NGC3245} and~\ref{Panel_NGC4486}, respectively). 
  
  The differences in the H$\alpha$ profiles are important, but also there are dramatic variations for the ratios of the secondary with respect to the narrow component. The most extreme case is NGC 4594 (see Fig.~\ref{Panel_NGC4594} and Fig.~\ref{BPTs}), in which the H$\alpha$ contribution to the total flux of the narrow component is almost negligible in comparison to the contribution of the secondary and the broad component. However, the secondary component shows a very similar line ratios than for the Palomar spectrum. 

  In the top panel of Fig.~\ref{BPTs}, considering the [S\,II]/H$\alpha$ ratio, the great majority of points for ground-based data are located around the transition region (i.e. log([S\,II]/H$\alpha$)$\sim$0.0). In these cases, the information of the [O\,III]/H$\beta$ ratio is required to further constraint the source of the ionization field. This is true for all objects except for NGC 4736, which is the only galaxy for which both [N\,II]/H$\alpha$ and [S\,II]/H$\alpha$ line ratios of the narrow component are consistent with being produced by the AGN, as mentioned above.
  Our classification, based on the [S\,II]/H$\alpha$ ratio, is consistent with the presence of a very broad component (BLR-originated) visible in its \textit{HST} spectrum (Fig.~\ref{Panel_NGC4736}). 
  
  On the other hand, for the classification based on [N\,II]/H$\alpha$ ratio, the majority of objects (7 out of 8) fall within the AGN region and only one is consistent with being ionized by a SF region. 

\section{Summary and conclusions}
\label{summary_conclusions}

In this work we present a kinematic analysis and the study of ionization mechanisms of the gas of 9 type-2 LINERs with both space- and ground-based spectroscopy obtained from the \textit{HST}/STIS archive and from the Palomar/Double Spectrograph \citep{Ho1995}, respectively. 

The main objective is the detection of the existence of a BLR in these galaxies by analyzing its spectral emission line features and the possible presence of outflows. Moreover, we addressed the line ratios of the emission lines to study the dominant ionization mechanism for these objects.

We used the [S\,II] and [O\,I] lines as templates to model the other forbidden lines ([N\,II]) and narrow H$\alpha$, and considered the addition of a broad component under H$\alpha$ when necessary. We used a maximum of 3 Gaussians component named narrow, secondary and broad component. Our approach allowed us to explore a larger number of scenarios for modeling emission lines with respect to previous work. These scenarios include possible NLR-statification and non-rotational motions being present, and never explored, for type-2 LINER-nuclei (see C18 for type-1 LINERs).

The narrow component is likely to be associated with the rotation from the galaxy disk, whereas the secondary component,of intermediate width, could be related to both rotational or non-rotational motions (as outflows). The broad component of the Balmer lines is always associated to the BLR of the AGN.

The secondary component has been detected for 4 LINERs in the \textit{HST} spectra and for 3 in the Palomar data. For two objects from the \textit{HST} analysis (i.e. NGC 4486 and NGC 4594), this component is found to suggestive of non rotational motions such as outflows. The line width of these components is similar to those found for outflows in type-1.9 LINERs by \citet{Cazzoli2018}. 

The detection of the BLR in LINERs is crucial to reveal its true AGN nature. \citet{Ho1997} optically classified 8 of the galaxies as type-2 objects, because their line fitting did not require a very broad component BLR-originated to reproduce the H$\alpha$-[N\,II] profile. With our analysis (following C18) of the same Palomar-spectra, we have found that for 2 out of 8 LINERs, a broad component is indeed necessary to reproduce emission lines. For the space-based data a very broad component is needed in 6 out of the 9 analyzed objects. 

The only detection of this component in the Palomar data but not confirmed with the \textit{HST} is for NGC 4486. The results for this component are not well constrained, thus we consider the BLR component detection as ambiguous. 
NGC 4594 is the only galaxy in the sample for which a broad component is needed in both space- and ground-based spectra, indicating that it would be better classified as as a type-1 LINER. 

Three of the BLR detections in the \textit{HST} data (NGC 3245, NGC 4594 and NGC 4736) were already reported by \citet{Constantin2015}. Our measurements of both FWHM and flux contribution of the broad component with respect to the total flux of the H$\alpha$-[N\,II] complex agree with \citet{Constantin2015}.

The detection of a very broad component is favored with the \textit{HST} data as the spatial resolution is 10 times better than for the Palomar data, decreasing the amount of contaminant starlight from the host galaxy and preventing the possible dilution of the broad component. The spatial resolution could be also relevant for the line diagnostic diagrams, as the calculated line ratios vary significantly from ground- to space-based spectra.

\begin{acknowledgements}
Authors acknowledge the anonymous referee for his/her constructive comments to improve the paper.
We acknowledge financial support by the Spanish Ministerio de Ciencia, Innovaci{\'o}n y Universidades (MCIU) under the grant AYA2016-76682-C3 and from the State Agency for Research of the Spanish MCIU through the "Center of Excellence Severo Ochoa" award to the Instituto de Astrof{\'i}sica de Andaluc{\'i}a (SEV-2017-0709). LHM acknowledge finantial support under the grant BES-2017-082471. We acknowledge O. Reyes-Amador for his useful help on this work. 

This research has made use of the NASA/IPAC Extragalactic Database (NED), which is operated by the Jet Propulsion Laboratory, California Institute of Technology, under contract with the National Aeronautics and Space Administration.
We acknowledge the usage of the HyperLeda database (http://leda.univ-lyon1.fr).
This work has made extensive use of IRAF and Python, particularly with \textsc{astropy}  (\nolinkurl{http://www.astropy.org}) \citep{astropy:2013, astropy:2018}, \textsc{matplotlib} \citep{Hunter:2007}, \textsc{numpy} and \textsc{lmfit}.
\end{acknowledgements}

%
   \bibliographystyle{aa} 
   \bibliography{aa_LINERs2_v1_referee.bbl} 

%






   
%
%
%


\begin{appendix} 
\section{Stellar modeling and subtraction for Palomar spectra}
\label{appendix_B}

The starlight modeling and subtraction to the ground-based spectra is presented here for the 8 available LINERs. The stellar continuum in the observed spectra was modeled with the \textsc{ppxf} method as in \citet{Cazzoli2018} (Sect.~\ref{analysis}).

\begin{figure*}
\centering
   \includegraphics[clip,trim=0.5cm 19.5cm 0.5cm 3.3cm,width=\textwidth]{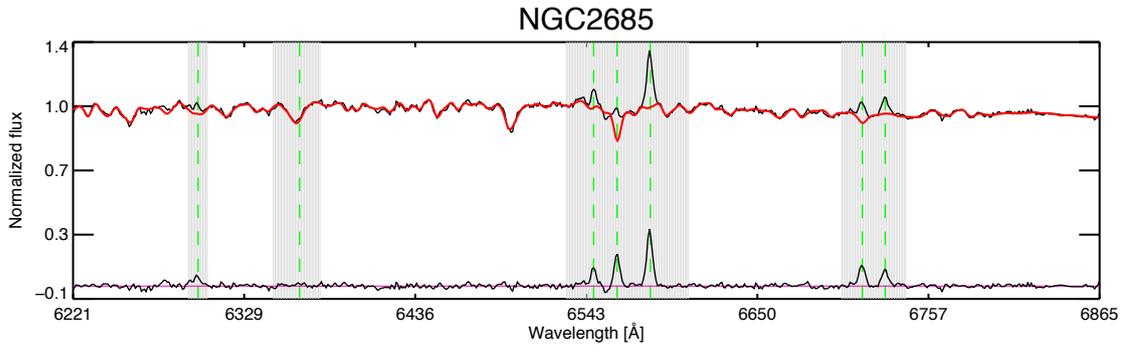}
   \caption{Optical ground-based Palomar spectrum of NGC 2685. 
            The original data is shown in black; the red line indicates the modelled stellar spectrum that matches the observed stellar continuum. The spectrum used for the kinematical decomposition (i.e. data $-$ model) is also presented. Green dashed lines show the rest-frame position of the [O\,I], [N\,II]-H$\alpha$ and [S\,II] lines. Grey bands indicate the regions masked to perform the stellar modeling. The magenta line indicates the zero-level as reference.} 
   \label{StSbs_NGC2685}
\end{figure*}

\begin{figure*}
\centering
   \includegraphics[clip,trim=0.5cm 19.5cm 0.5cm 3.3cm,width=\textwidth]{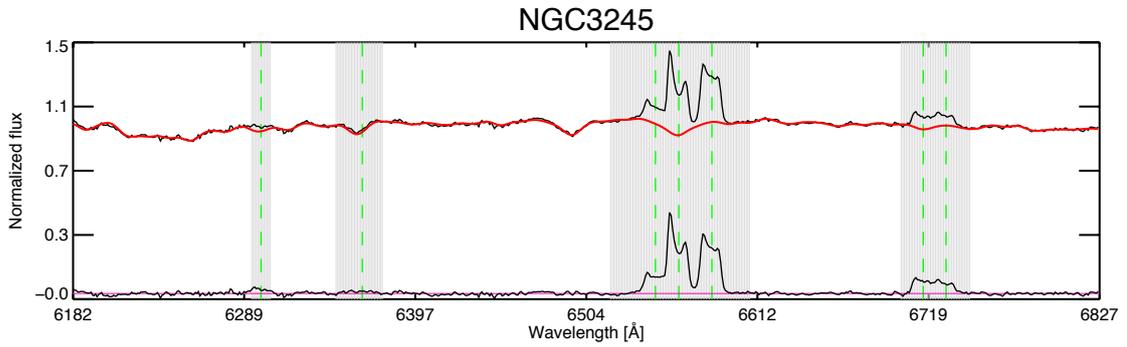}
   \caption{Same as Fig.~\ref{StSbs_NGC2685} but for NGC 3245.}
\end{figure*}

\begin{figure*}
\centering
   \includegraphics[clip,trim=0.5cm 19.5cm 0.5cm 3.3cm,width=\textwidth]{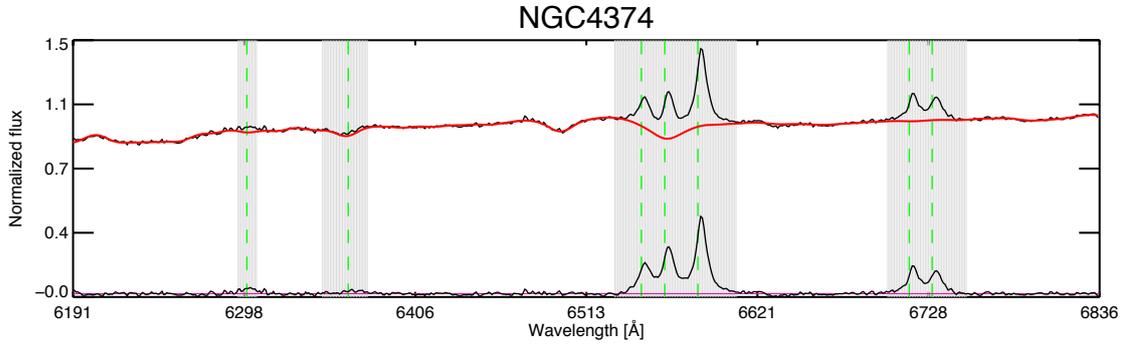}
   \caption{Same as Fig.~\ref{StSbs_NGC2685} but for NGC 4374.}
\end{figure*}

\begin{figure*}
\centering
   \includegraphics[clip,trim=0.5cm 19.5cm 0.5cm 3.3cm,width=\textwidth]{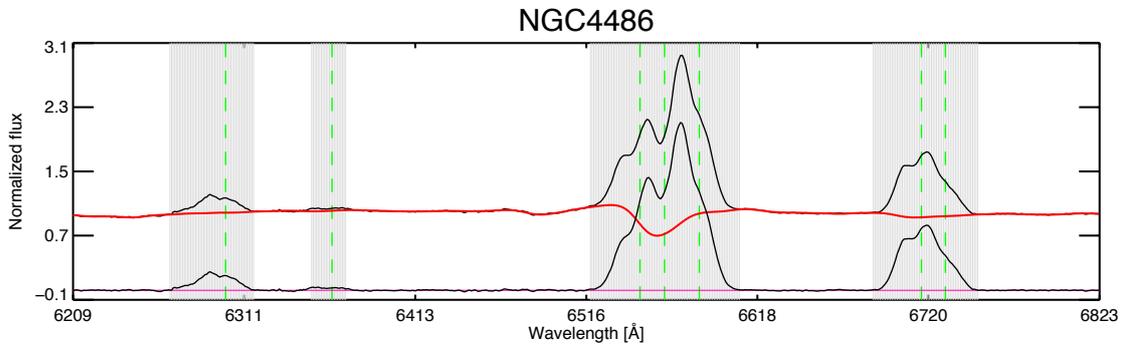}
   \caption{Same as Fig.~\ref{StSbs_NGC2685} but for NGC 4486.}
\end{figure*}

\begin{figure*}
\centering
   \includegraphics[clip,trim=0.5cm 19.5cm 0.5cm 3.3cm,width=\textwidth]{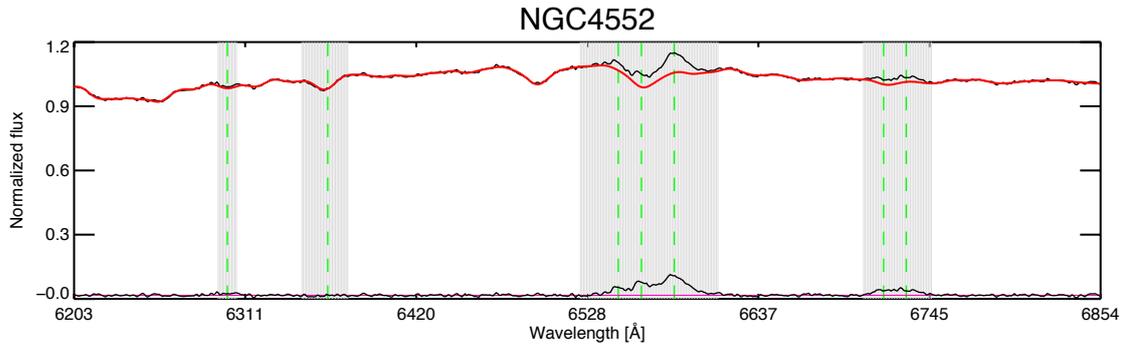}
   \caption{Same as Fig.~\ref{StSbs_NGC2685} but for NGC 4552.}
\end{figure*}

\begin{figure*}
\centering
   \includegraphics[clip,trim=0.5cm 19.5cm 0.5cm 3.3cm,width=\textwidth]{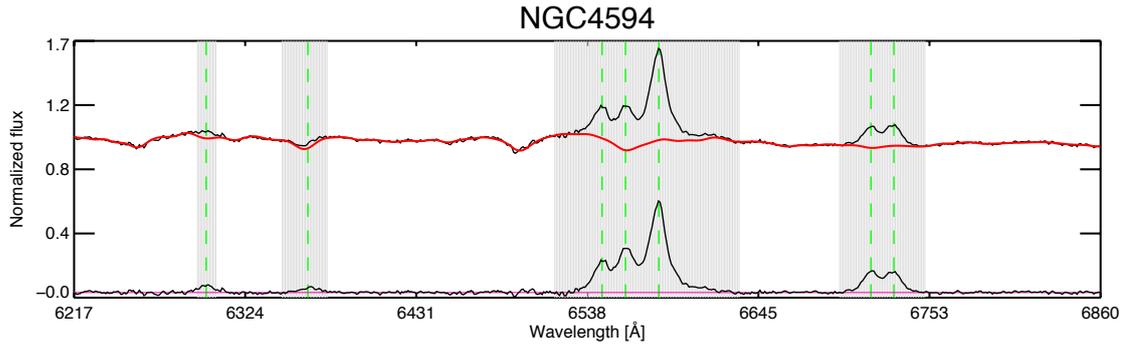}
   \caption{Same as Fig.~\ref{StSbs_NGC2685} but for NGC 4594.}
\end{figure*}

\begin{figure*}
\centering
   \includegraphics[clip,trim=0.5cm 19.5cm 0.5cm 3.3cm,width=\textwidth]{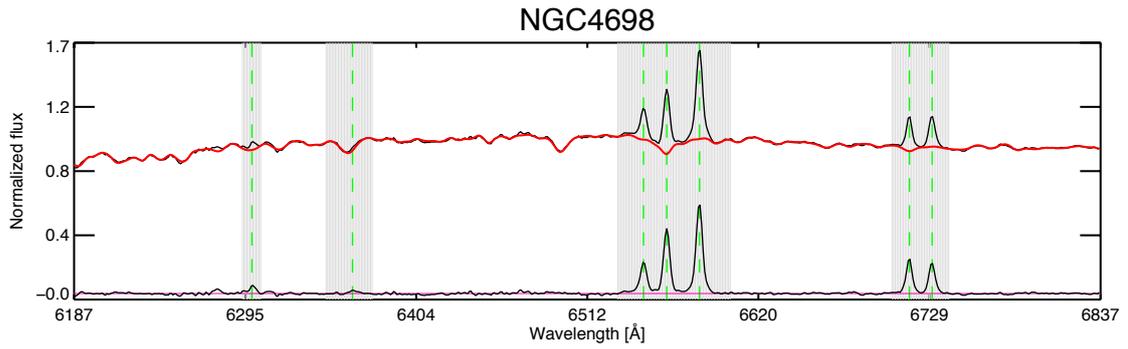}
   \caption{Same as Fig.~\ref{StSbs_NGC2685} but for NGC 4698.}
\end{figure*}

\begin{figure*}
\centering
   \includegraphics[clip,trim=0.5cm 19.5cm 0.5cm 3.3cm,width=\textwidth]{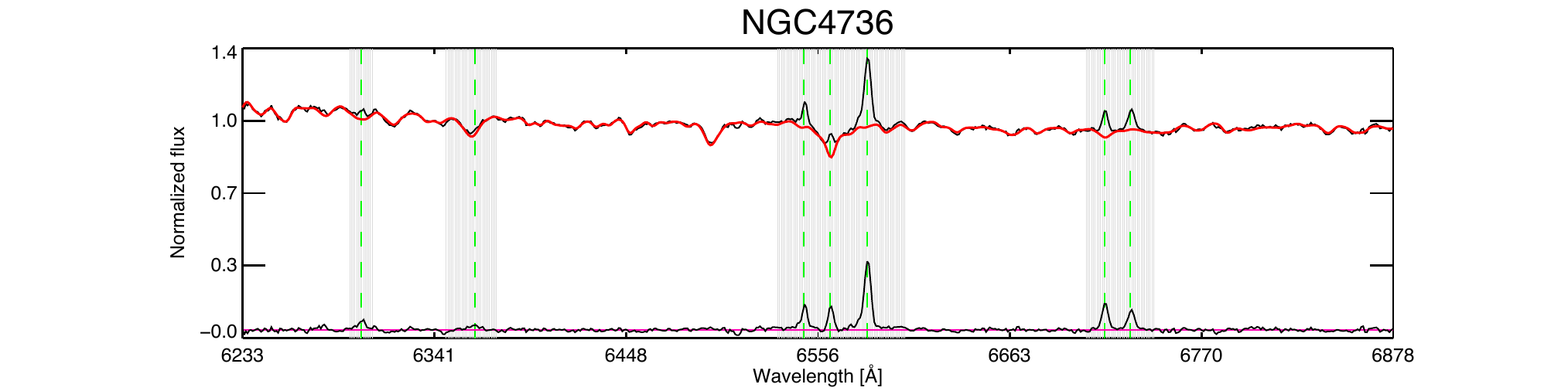}
   \caption{Same as Fig.~\ref{StSbs_NGC2685} but for NGC 4736.}
\end{figure*}

\section{Spectra modeling for each galaxy}
\label{appendix}

We present \textit{HST} images from the archive and the spectra analyzed in this manuscript for each LINER. They are arranged as follows.

\begin{itemize}
\item \textit{Top}: sharp divided \citep{Marquez1996,Marquez1999,Marquez2003} \textit{HST} image of the galaxy in the inner 10$^{''}$x10$^{''}$. The HST filters are summarized in Table~\ref{obslog}, column~9. Sharp-divided images are very useful to trace asymmetries in the light distribution, such as bars, spiral arms, dust lanes and rings. The method consists on dividing the original image by that obtained from the convolution with a median filter of 30 pixels. This technique allows the subtraction of the diffuse background emission facilitating the detection of small-scale variations and discuss the possible presence of both dust extinguished and more luminous regions \citep[e.g.][]{Masegosa2011}. 
The galaxy ID is indicated in the panel title. White and yellow continuous lines represent the slit used to obtain the \textit{HST}/STIS and Palomar spectrum (when available), respectively. 
The angular scale of 1\arcsec of the image in parsecs (see Table~\ref{sample} column~6) is indicated in the right-upper part of the image. The white dashed line indicates the PA of the major axis (see Tab.~\ref{sample}).
\item \textit{Middle}: line modeling of the \textit{HST}/STIS spectra (see Sect.~\ref{analysis} for details). We marked with different colors each Gaussian component (same colours mark the same kinematic components). Green lines represent the narrow component; when present, blue lines represent the secondary component and purple lines, the broad component. The red curve shows the total contribution of all the components. The continuum range selected to calculate the standard deviation is plotted in yellow. Residuals from the fits are in the middle/lower panels. We have included the residuals of the fit without the broad component for those objects in which this component was necessary (labeled as: \lq\textit{Res\,1}\rq) for comparison. When the two are plotted, the final fit residuals are labeled as: \lq\textit{Res \,2}\rq. In these panels, the orange lines indicate the 3$\sigma$ limits. If the [O\,I] lines are not present, their restframe position is indicated in gray in both spectrum and residual.  Wavelengths are restframe.
\item \textit{Bottom}: line modeling for the Palomar data (same description as in previous panel) after stellar continuum modeling and subtraction (see Sect.~\ref{analysis} and Appendix~\ref{appendix_B}). The flux is in arbitrary units.
\end{itemize}

\begin{figure}
\centering
   \includegraphics[width=.4\textwidth]{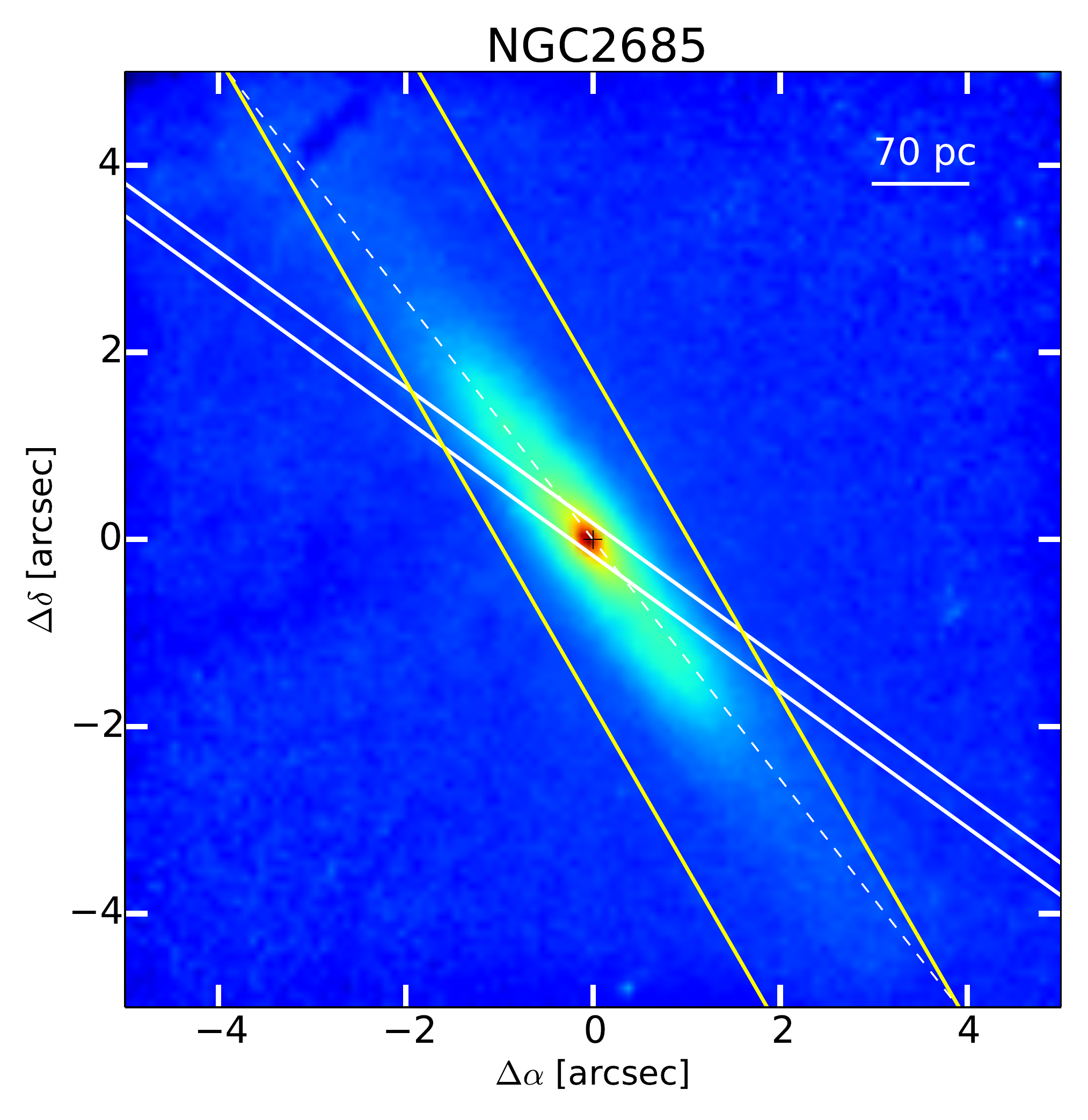}
   \includegraphics[width=\columnwidth]{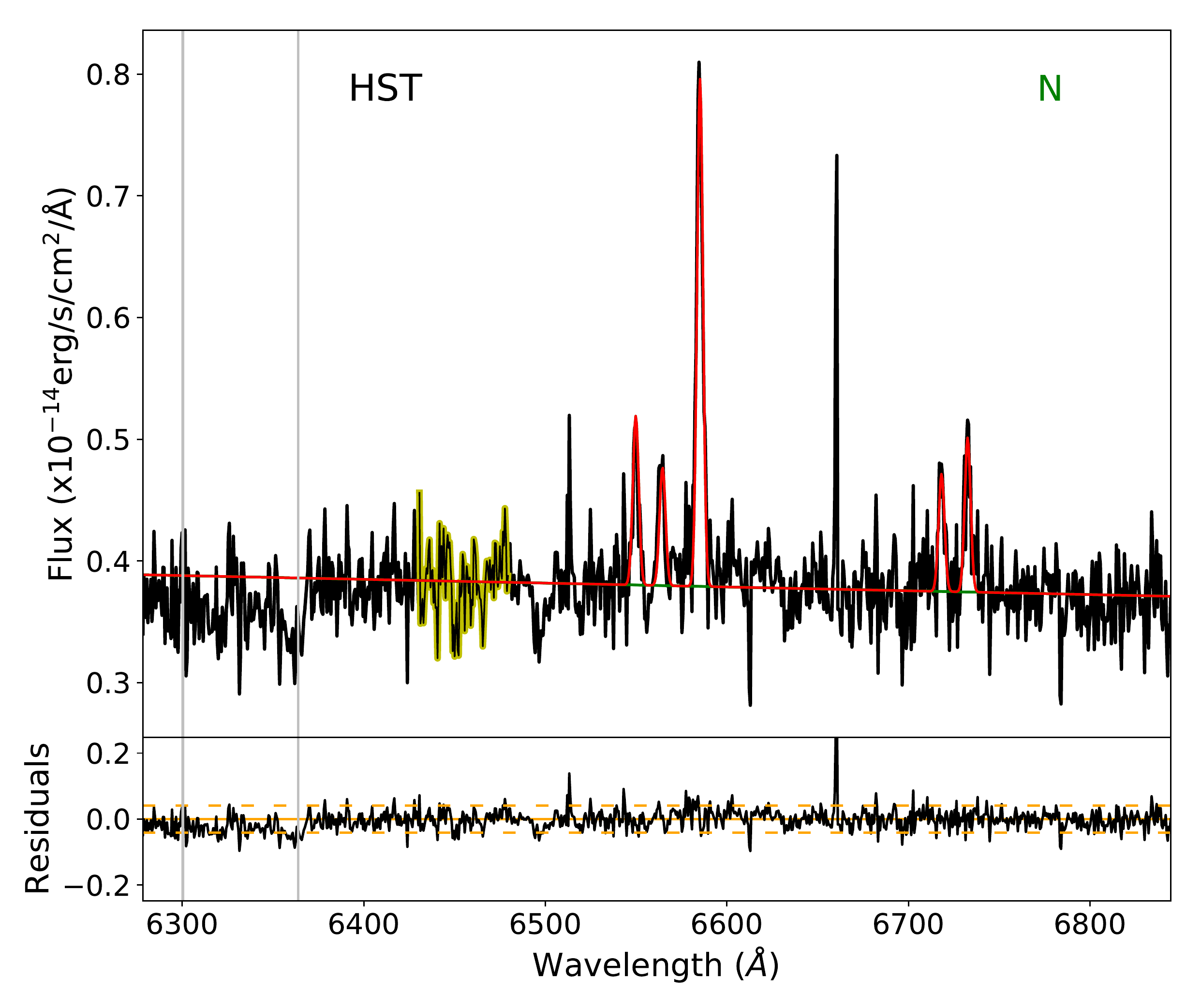} 
   \includegraphics[width=\columnwidth]{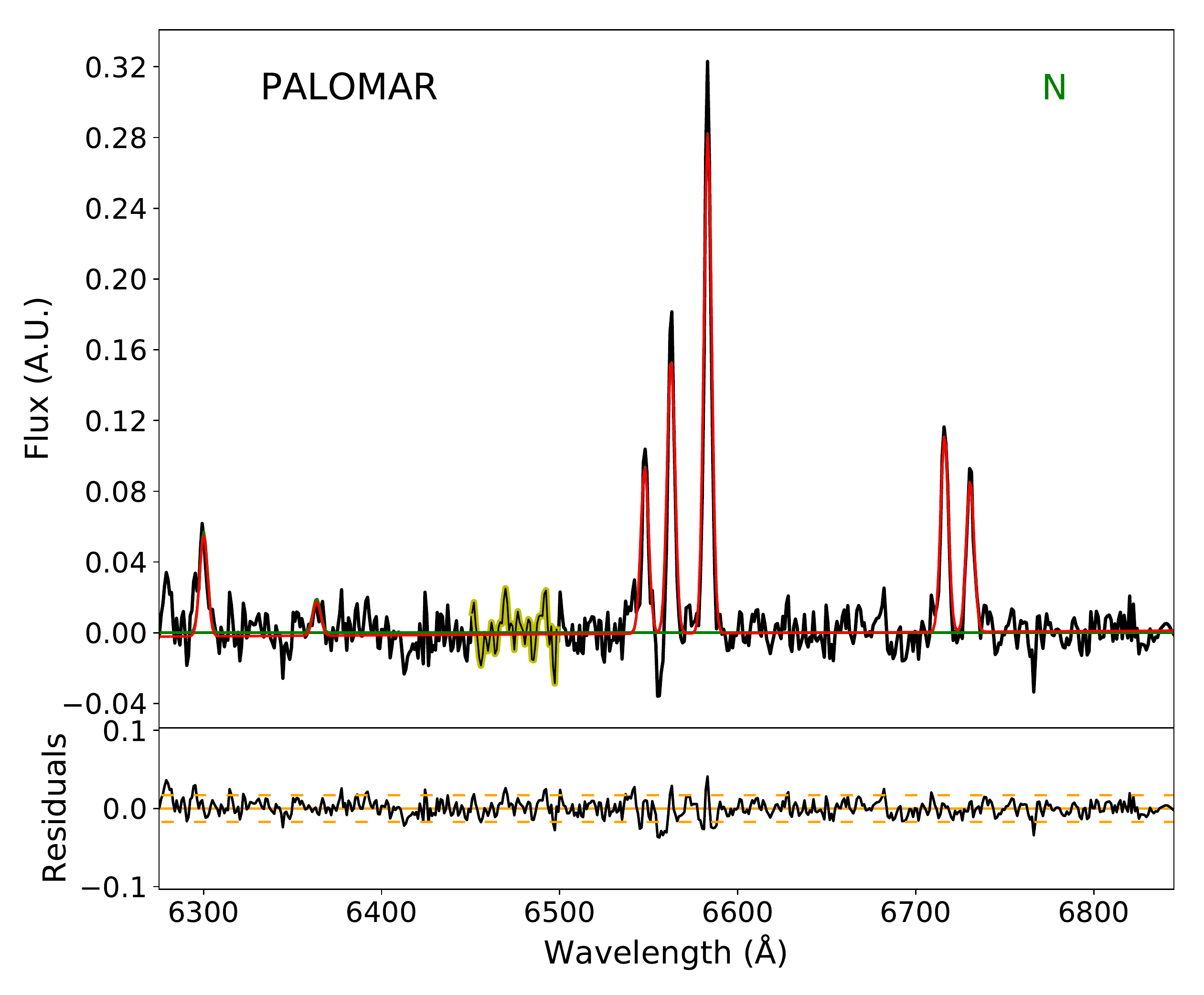}
 
 \caption{(General description in Appendix~\ref{appendix}). 
         NGC 2685: The H$\alpha$-[NII] lines are unblended in both spectra. [O\,I] lines are not present in the \textit{HST}/STIS spectrum, thus [S\,II] lines were used to model the other profiles. In the Palomar spectrum [O\,I] lines are visible, however no significant improvement was seen by using them to model the rest of the emission lines. All the lines are well reproduced with a single narrow Gaussian component.}
   \label{Panel_NGC2685} 		 		 
\end{figure}

\begin{figure}
\centering
   \includegraphics[width=.4\textwidth]{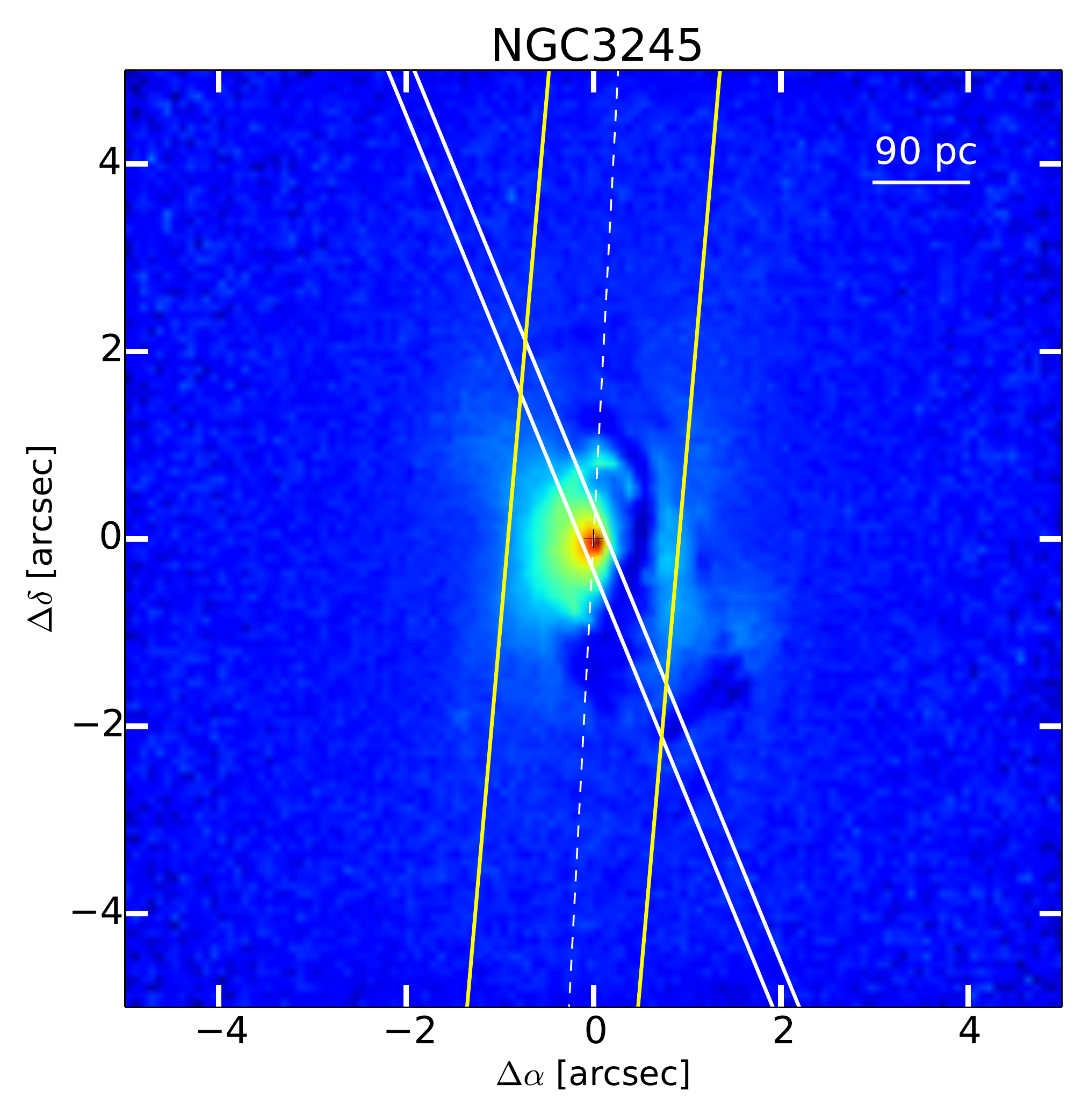} 
   \includegraphics[width=\columnwidth]{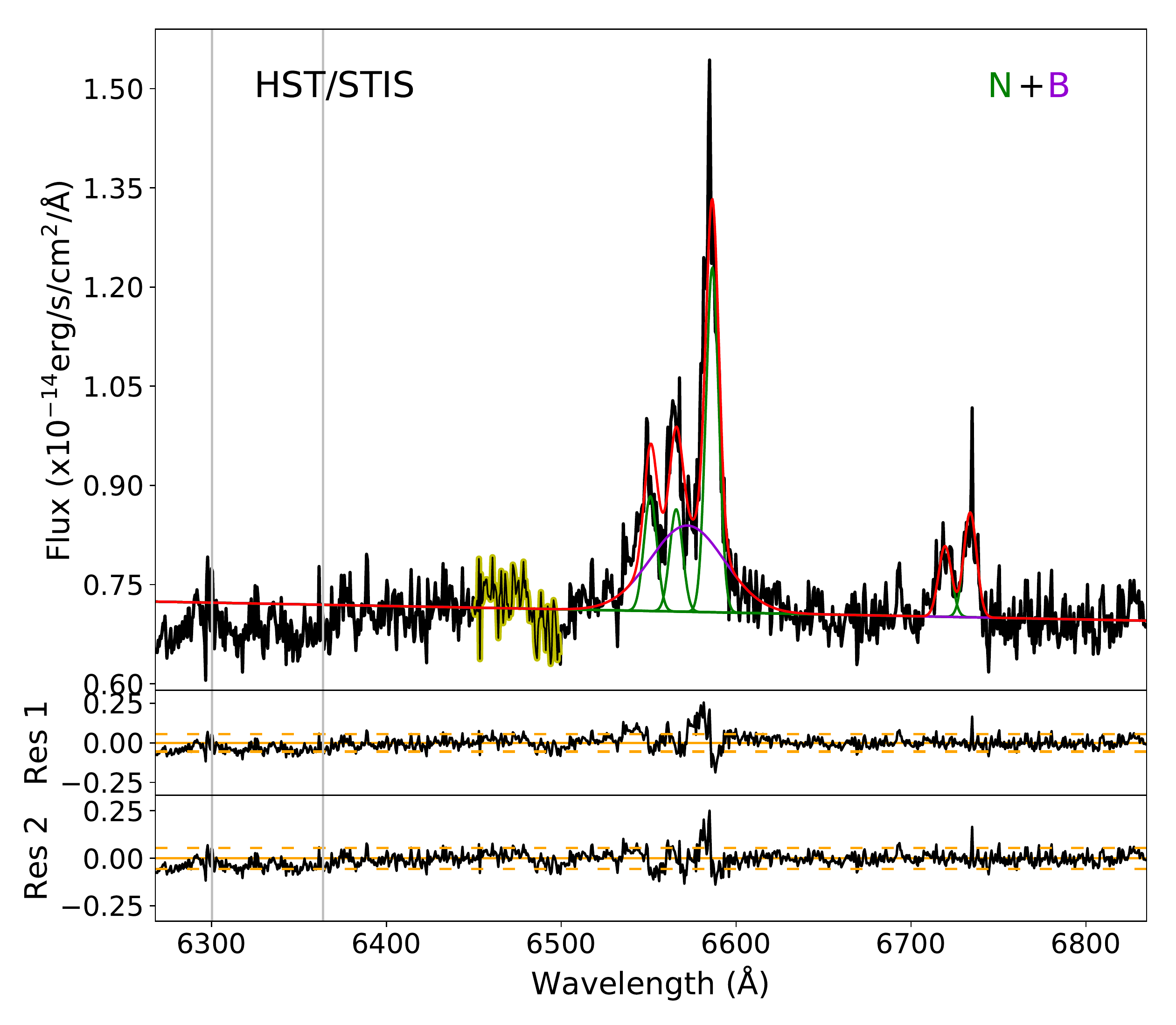} 
   \includegraphics[width=\columnwidth]{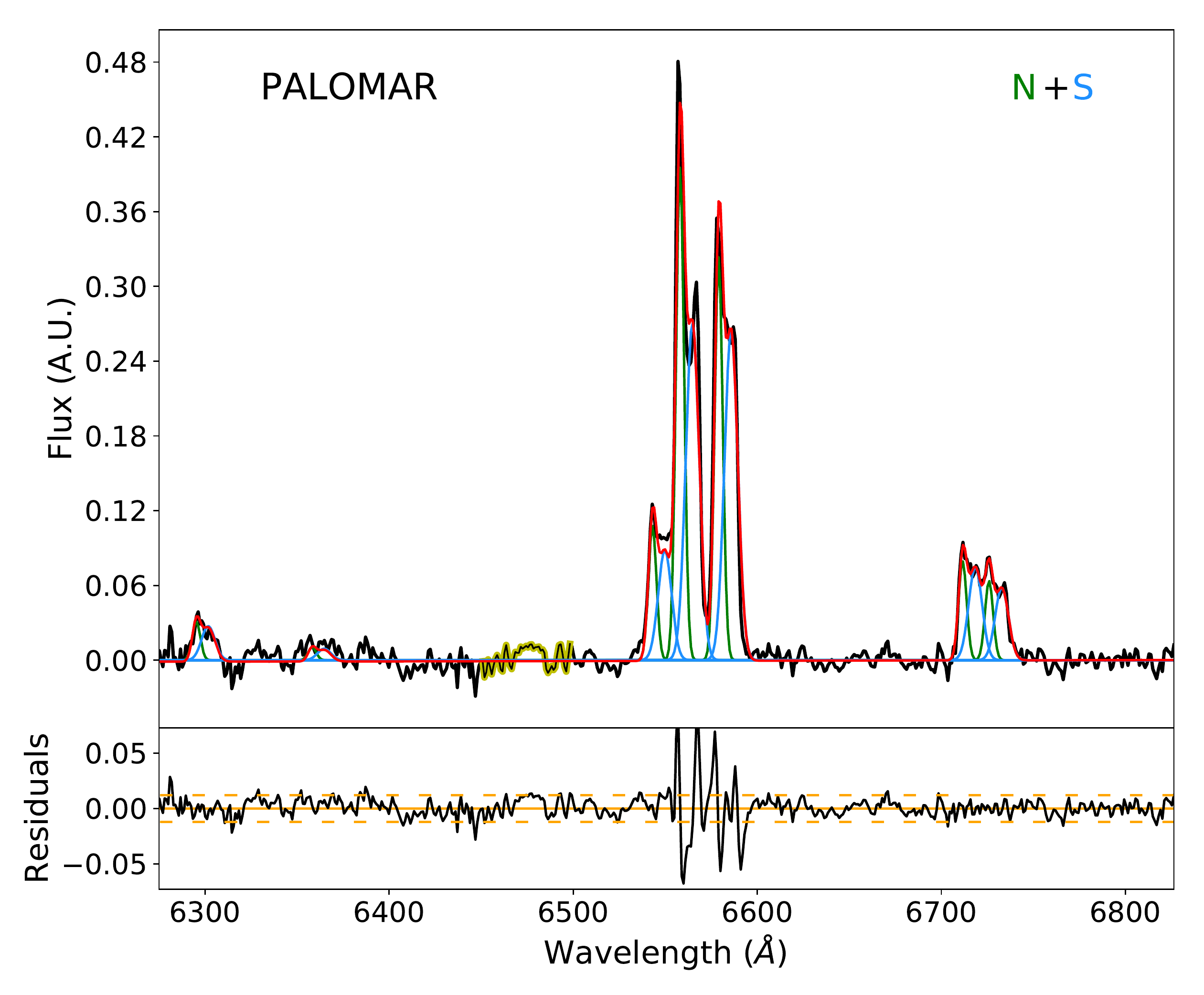} 

 \caption{(General description in Appendix~\ref{appendix}). 
         NGC 3245: For the \textit{HST}/STIS spectrum, the emission lines have been modeled with a single narrow Gaussian component. The residuals improved by adding a broad component to H$\alpha$. [O\,I] lines are not visible. In the Palomar spectrum, the double-peaked line profiles are modeled with two Gaussian components. As the [O\,I] line is rather faint, [S\,II] lines have been used for the modeling. 
         H$\alpha$ does not require any BLR-originated component.}
   \label{Panel_NGC3245} 		 		 
\end{figure}

\begin{figure}
\centering
   \includegraphics[width=.4\textwidth]{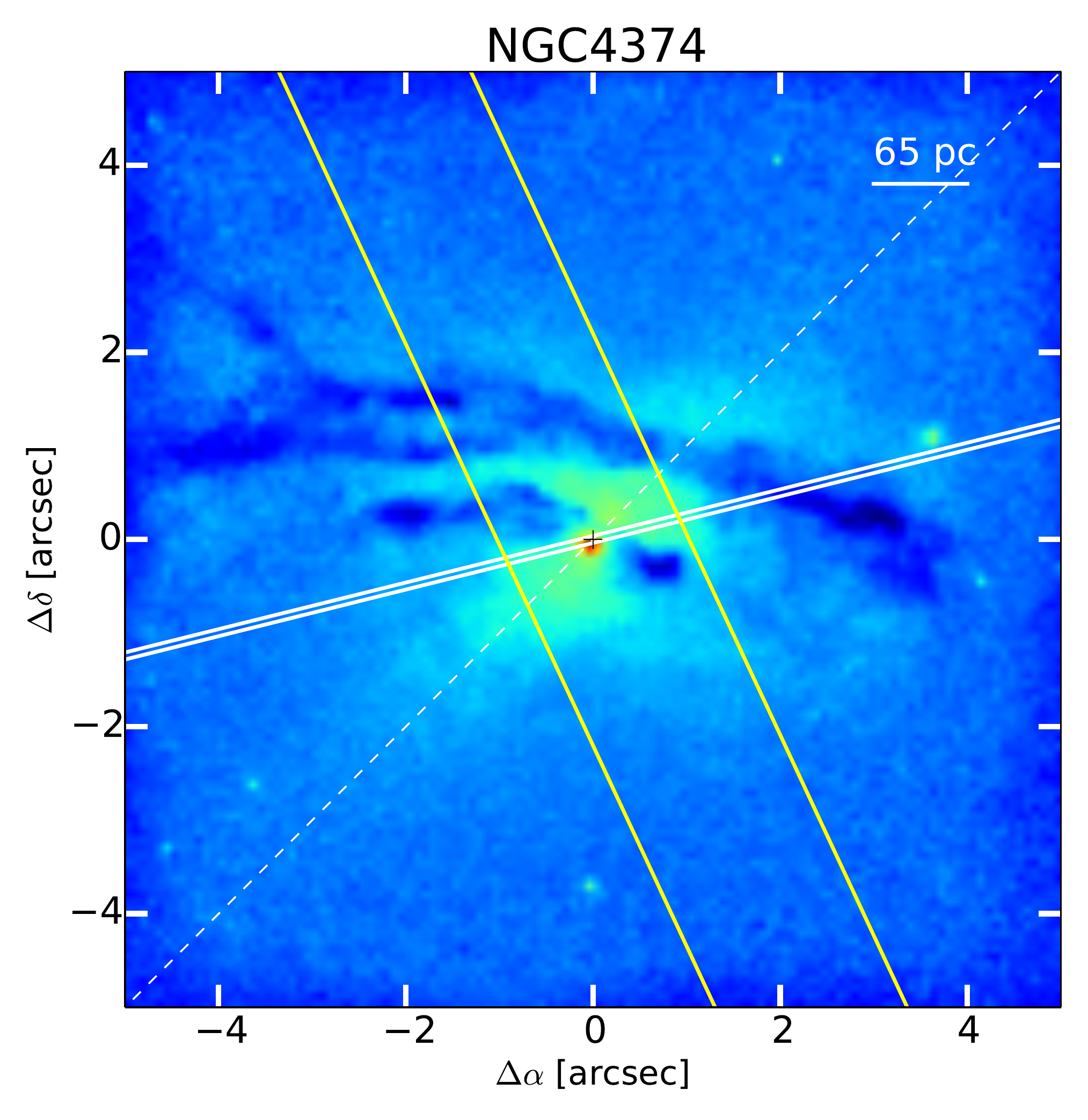}
   \includegraphics[width=.463\textwidth]{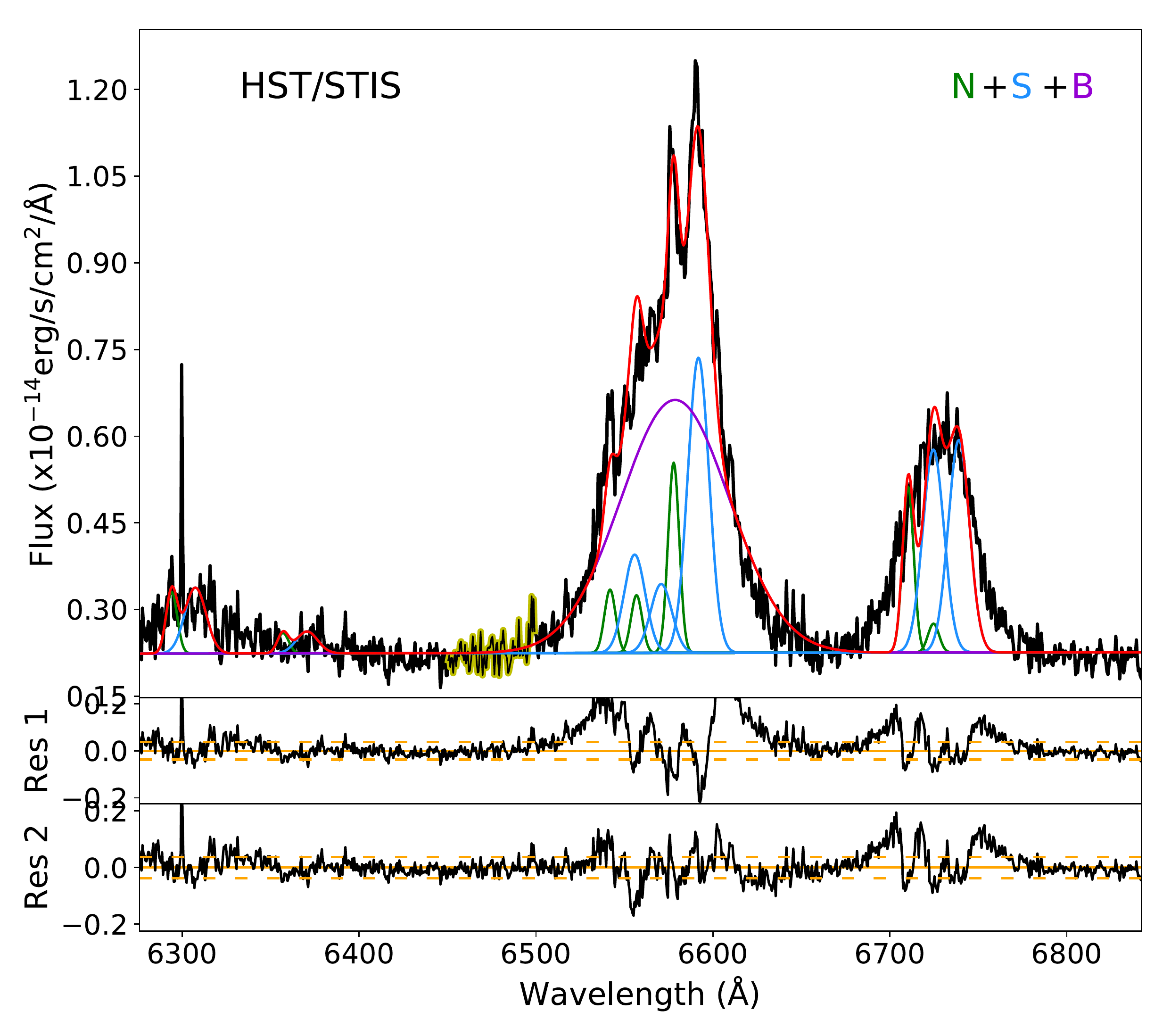} 
   \includegraphics[width=\columnwidth]{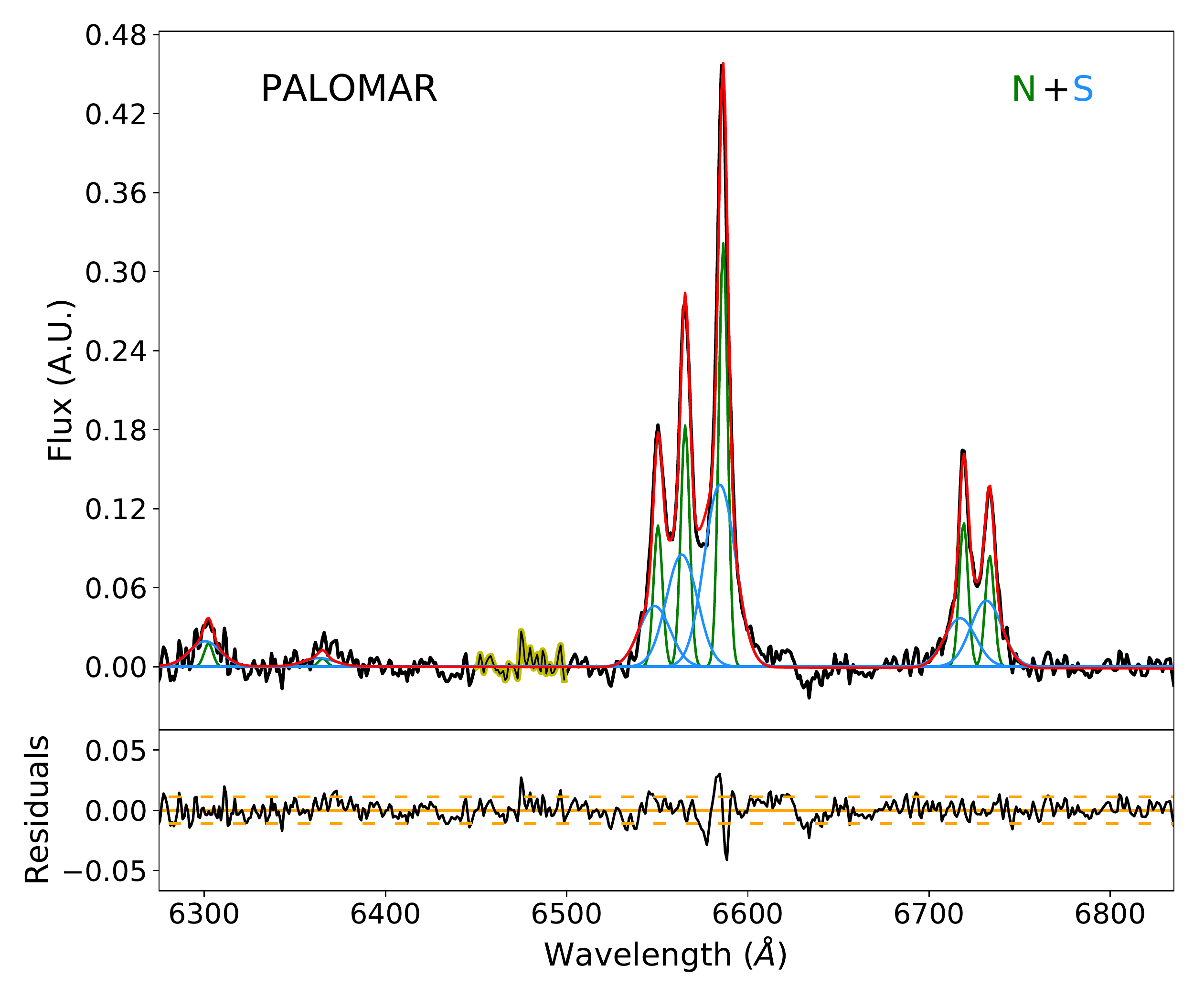} 
 
 \caption{(General description in Appendix~\ref{appendix}). 
         NGC 4374: In the \textit{HST}/STIS spectrum, [S\,II] lines are severely blended and only the [O\,I]$\lambda$6300\AA\, line is visible  (with low S/N). This complicated the modeling for the other lines, so the fit was performed considering the narrow peaks of [N\,II]-H$\alpha$ blend (see Sect.~\ref{analysis}), which results in high residuals under the [S\,II] lines. A broad component is needed in this decomposition.
         For the Palomar spectrum, all the emission lines are well modeled with two Gaussian components.}
   \label{Panel_NGC4374} 		 		 
\end{figure}

\begin{figure}
\centering
   \includegraphics[width=.4\textwidth]{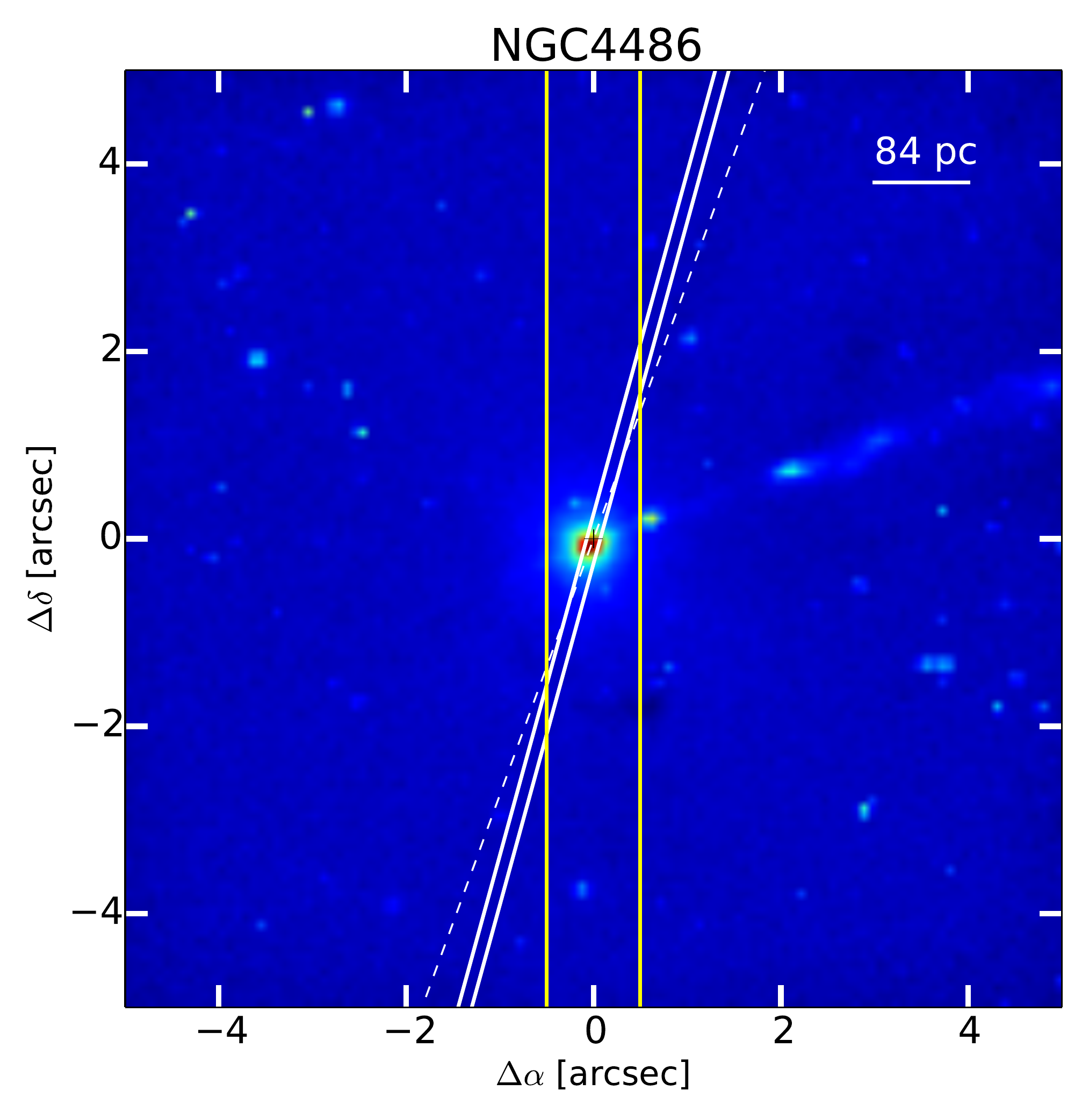} 
   \includegraphics[width=\columnwidth]{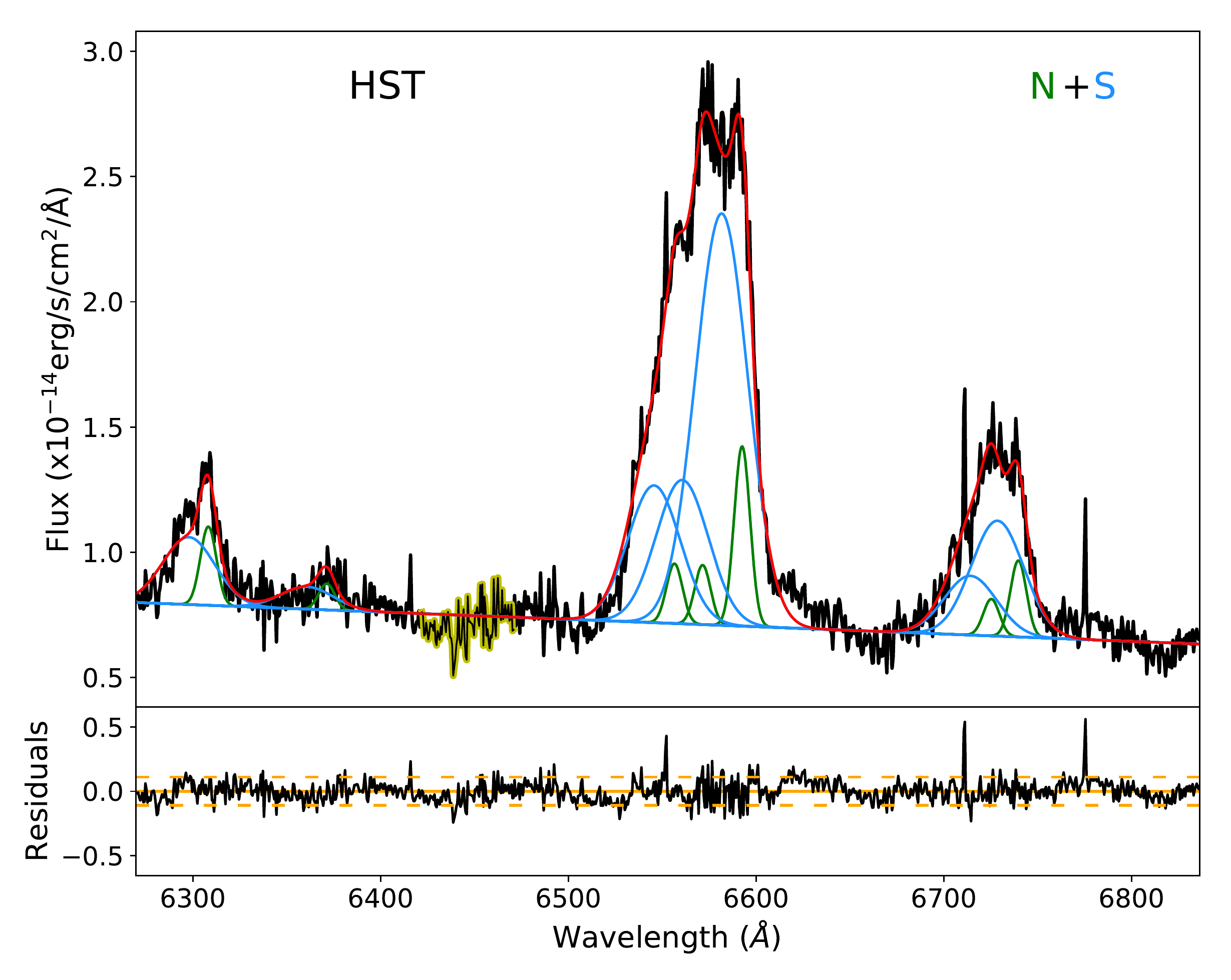} 
   \includegraphics[width=\columnwidth]{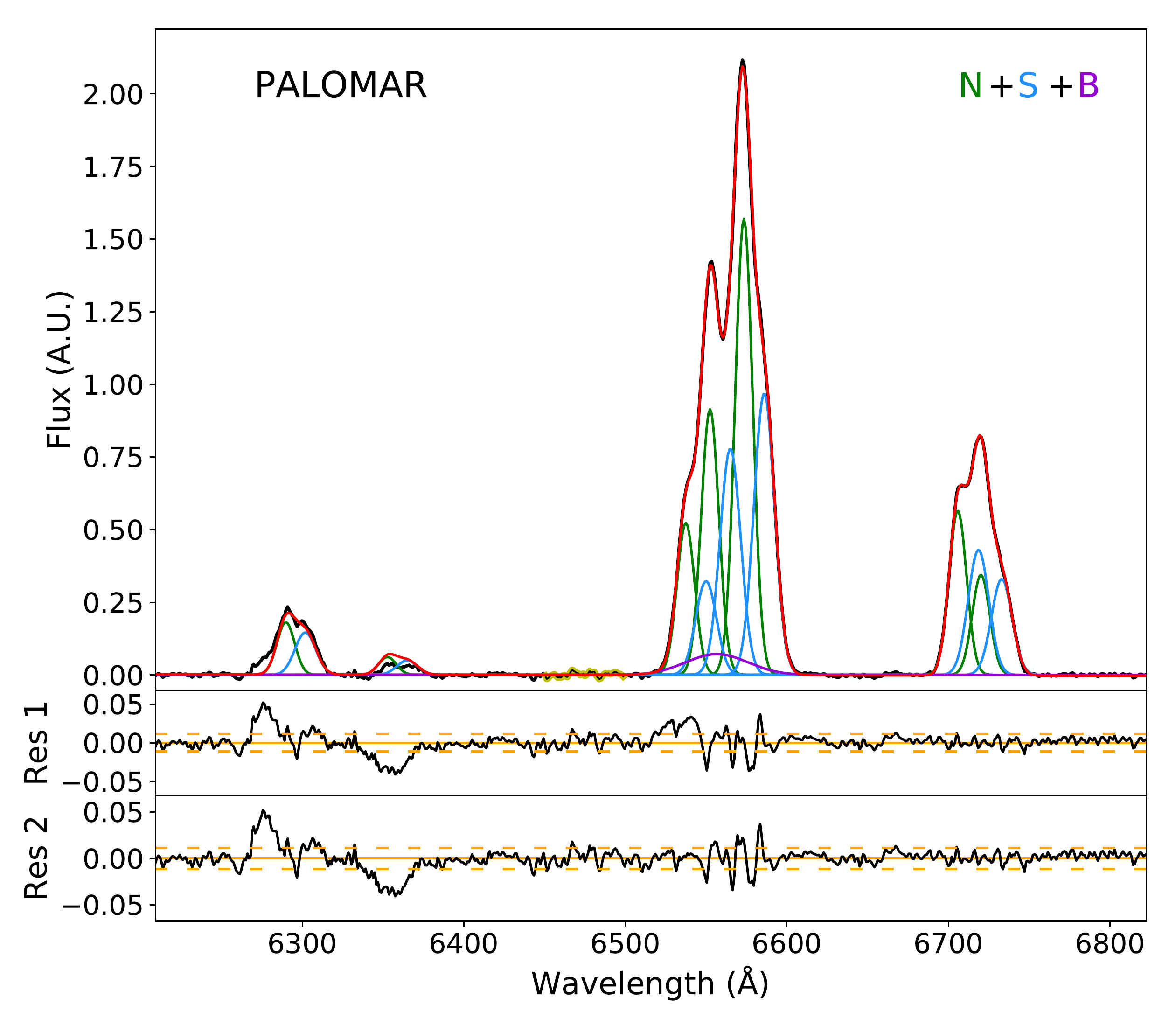} 

 \caption{(General description in Appendix~\ref{appendix}). 
         NGC 4486: In the \textit{HST}/STIS spectrum, two components are required to model the broad [O\,I] and [S\,II] emission. No broad component is needed to model the H$\alpha$ line profile. For the Palomar spectrum, two components have been used to reproduce the [S\,II] and [O\,I] lines, but a very faint broad component is used for H$\alpha$.}
   \label{Panel_NGC4486} 		 		 
\end{figure}

\begin{figure}
\centering
   \includegraphics[width=.4\textwidth]{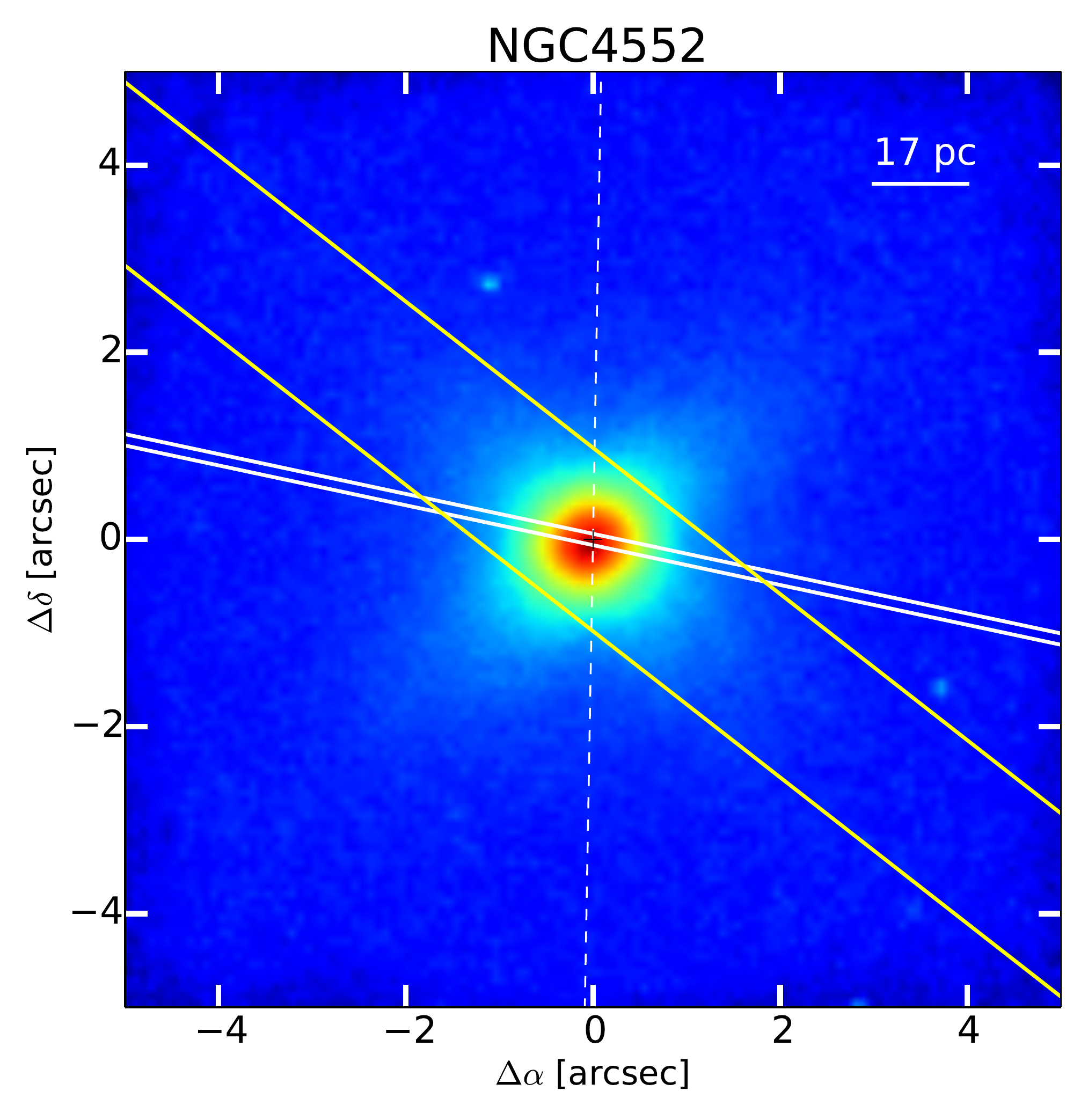} 
   \includegraphics[width=\columnwidth]{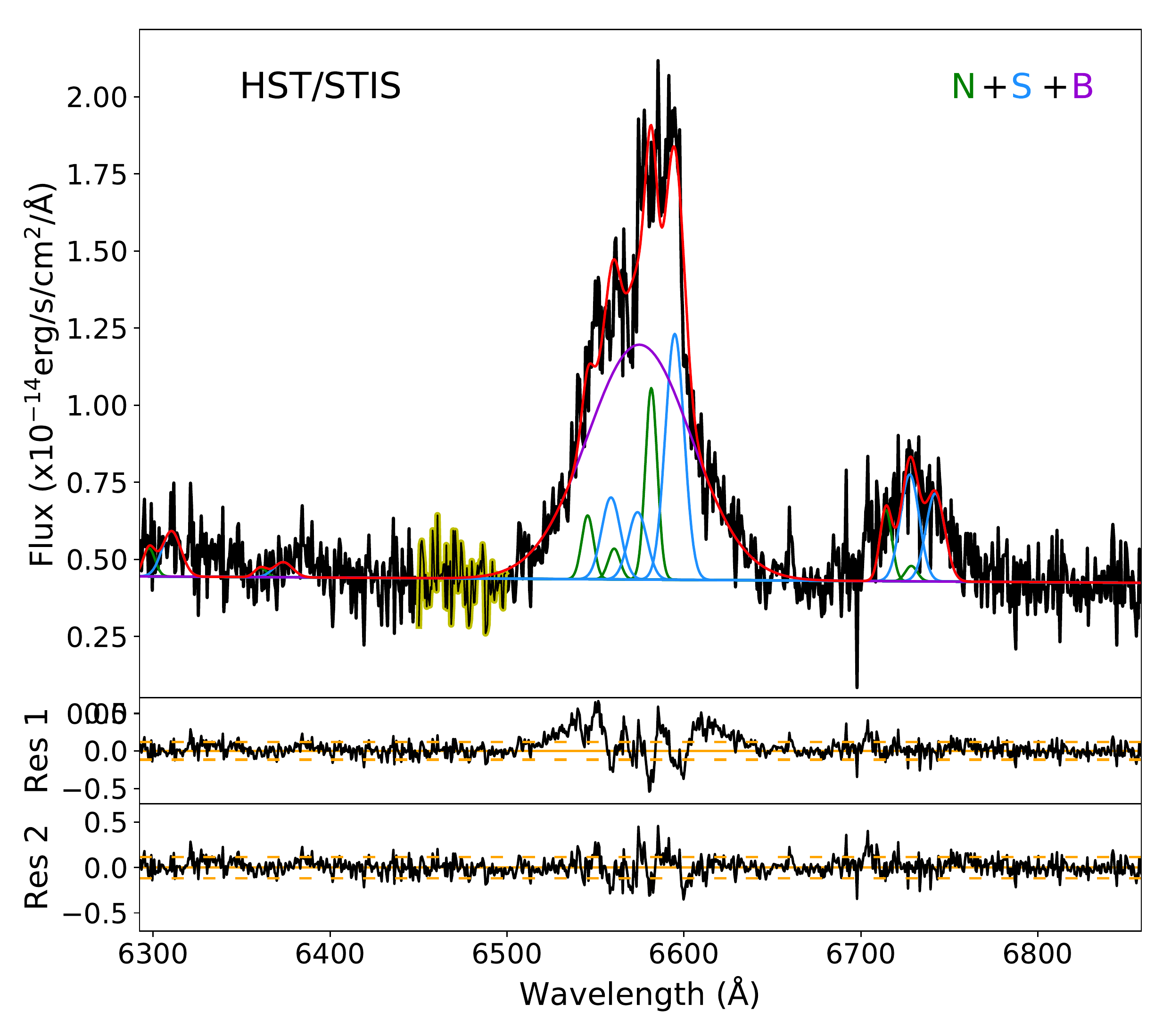}
   \includegraphics[width=\columnwidth]{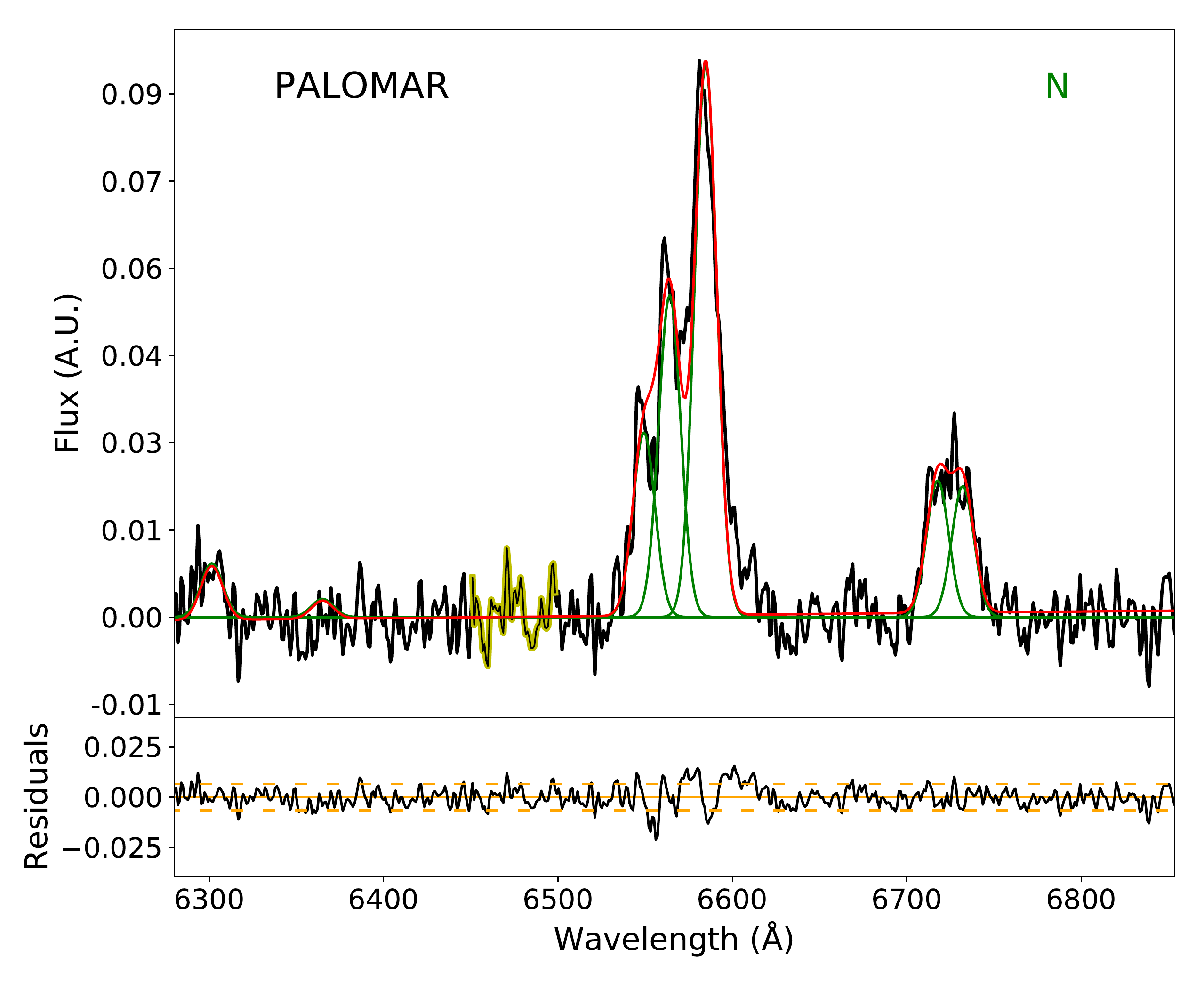}

 \caption{(General description in Appendix~\ref{appendix}). 
         NGC 4552: For the \textit{HST}/STIS spectrum, a second and a broad component were necessary to model emission lines. [O\,I] lines are very noisy and weak, thus they could not be used as a template for forbidden lines and narrow H$\alpha$. A broad component is not required to fit the H$\alpha$ profile. On the contrary, for the Palomar spectrum all the emission lines can be reproduced with only a single component.}
   \label{Panel_NGC4552} 		 		 
\end{figure}

\begin{figure}
\centering
   \includegraphics[width=.4\textwidth]{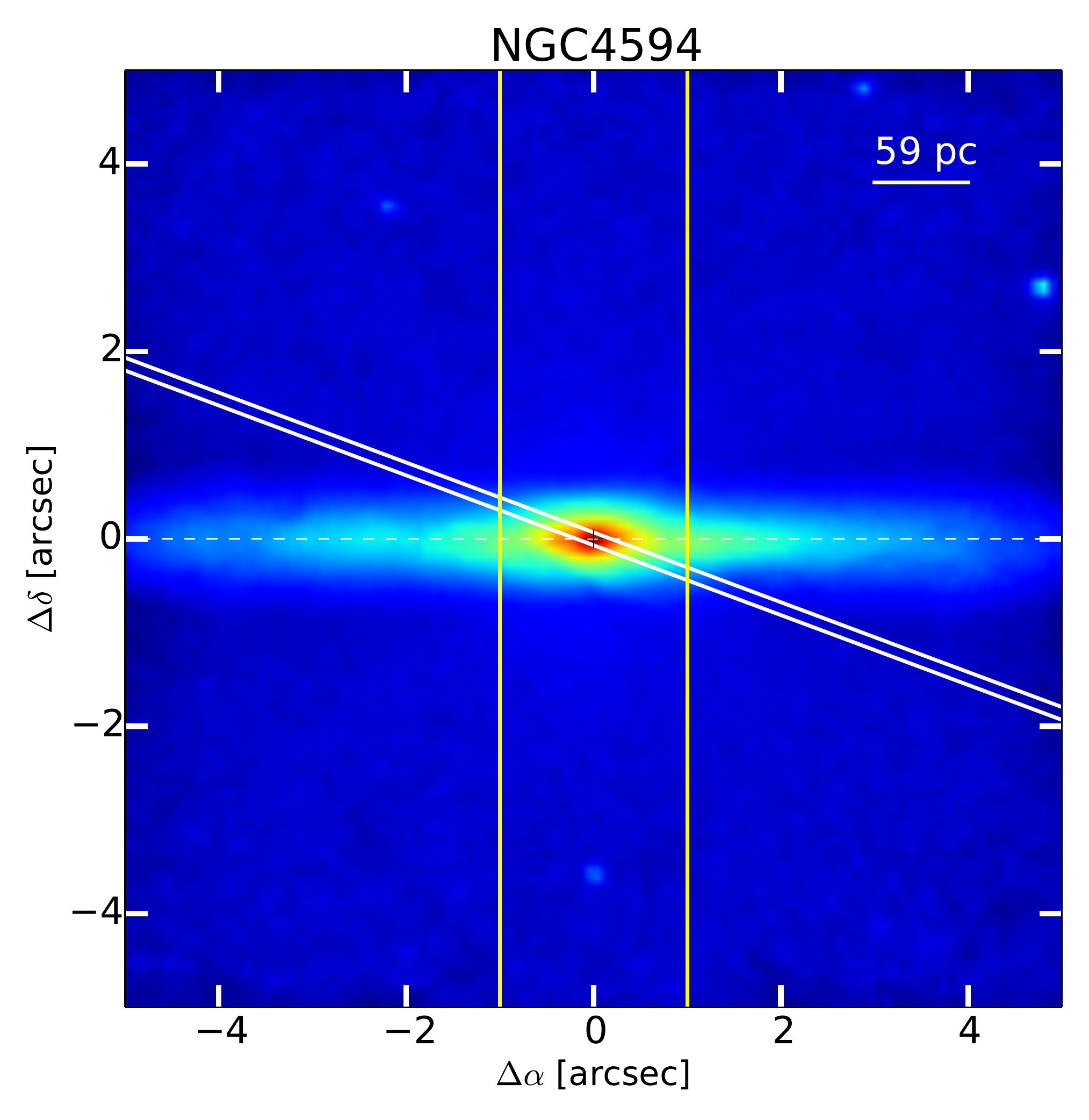}
   \includegraphics[width=.47\textwidth]{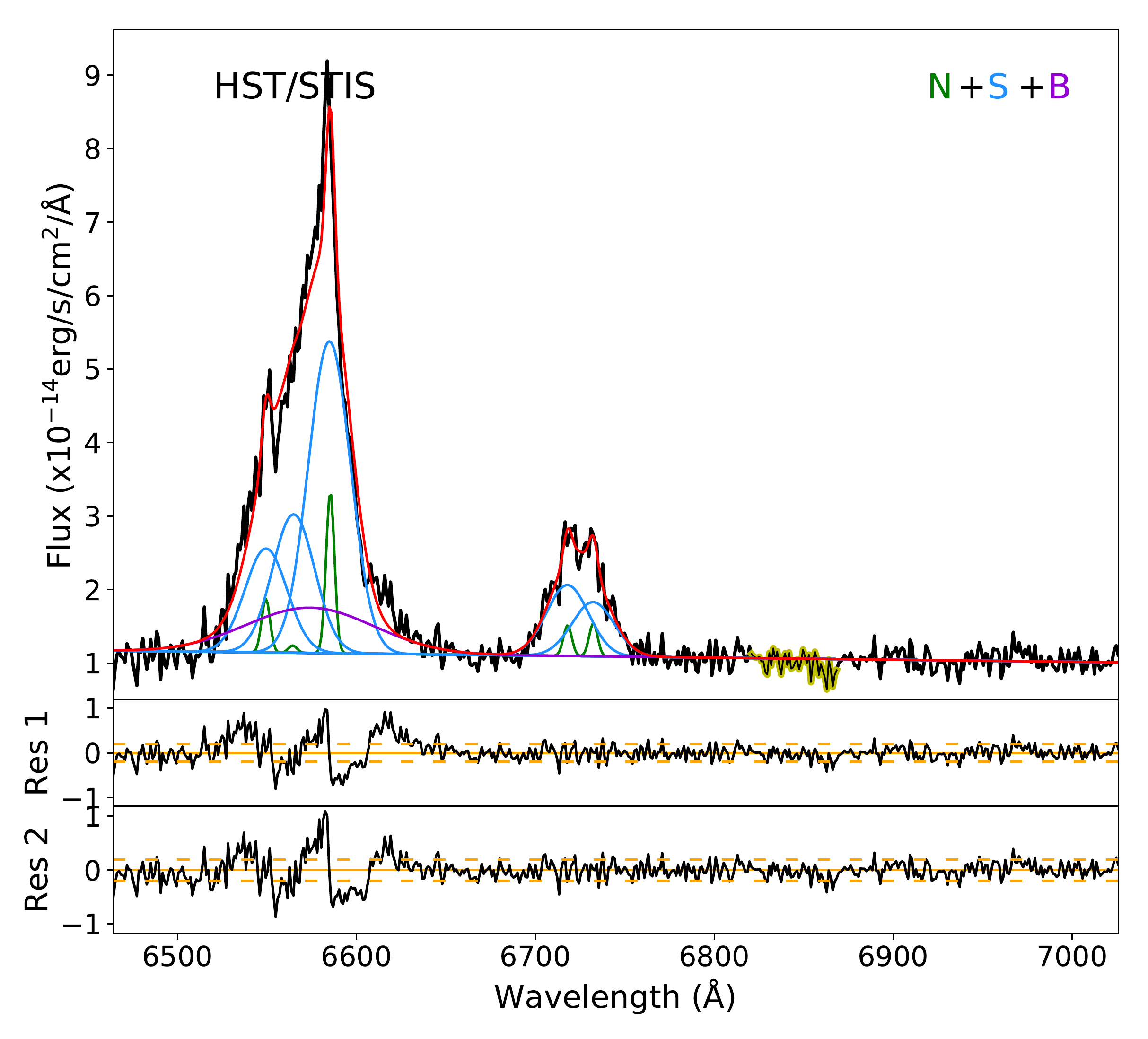}
   \includegraphics[width=.47\textwidth]{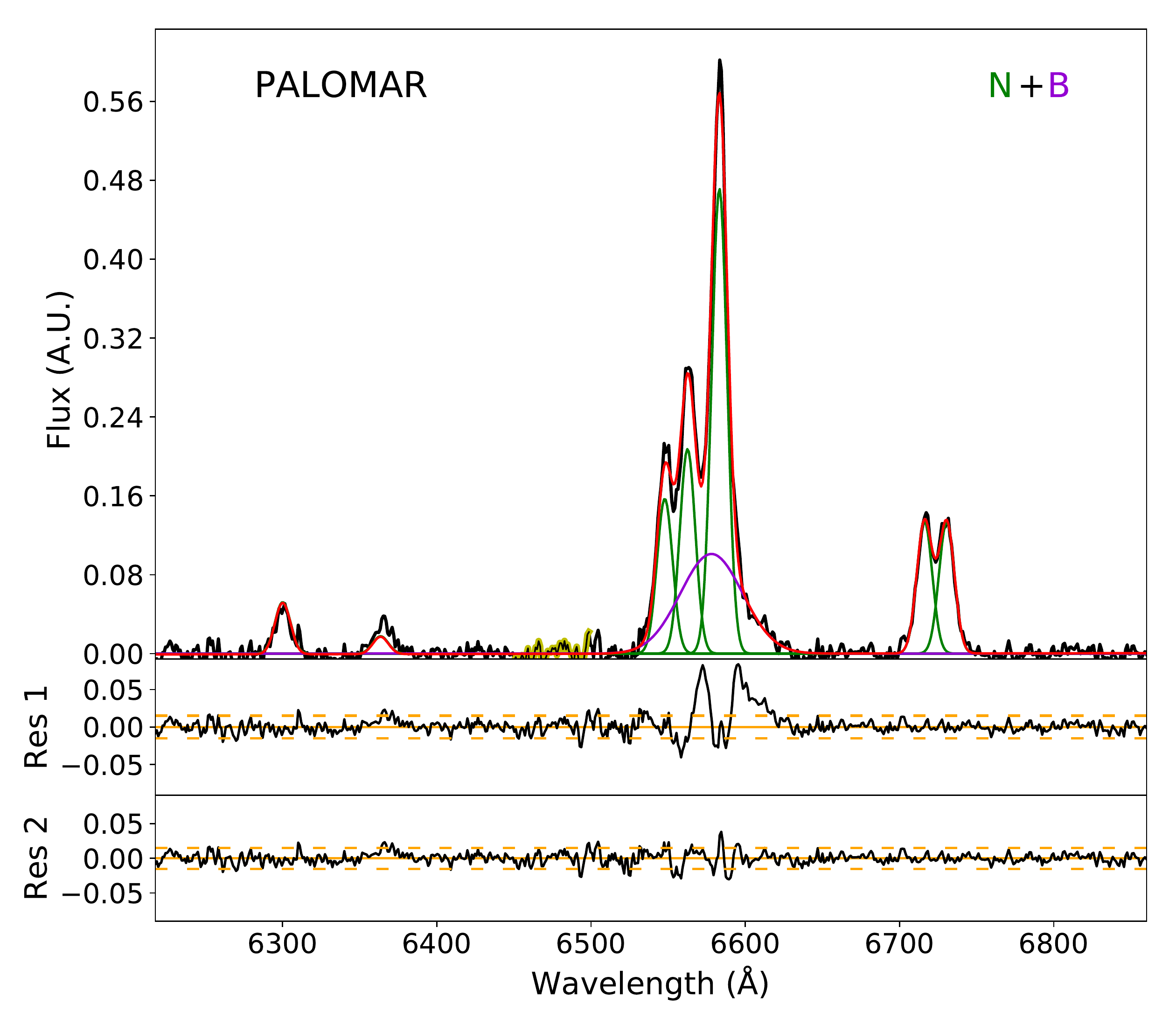}
 
 \caption{(General description in Appendix~\ref{appendix}). 
         NGC 4594: The wavelength coverage of \textit{HST}/STIS spectrum does not include [O\,I] lines. The [S\,II] lines have separate narrow profile peaks and a broader profile underneath. A broad component for H$\alpha$ is to improve the fit. This broad component is also needed for the Palomar spectrum, although a single narrow Gaussian is sufficient for the forbidden lines.}
   \label{Panel_NGC4594} 		 		 
\end{figure}

\begin{figure}
\centering
   \includegraphics[width=.4\textwidth]{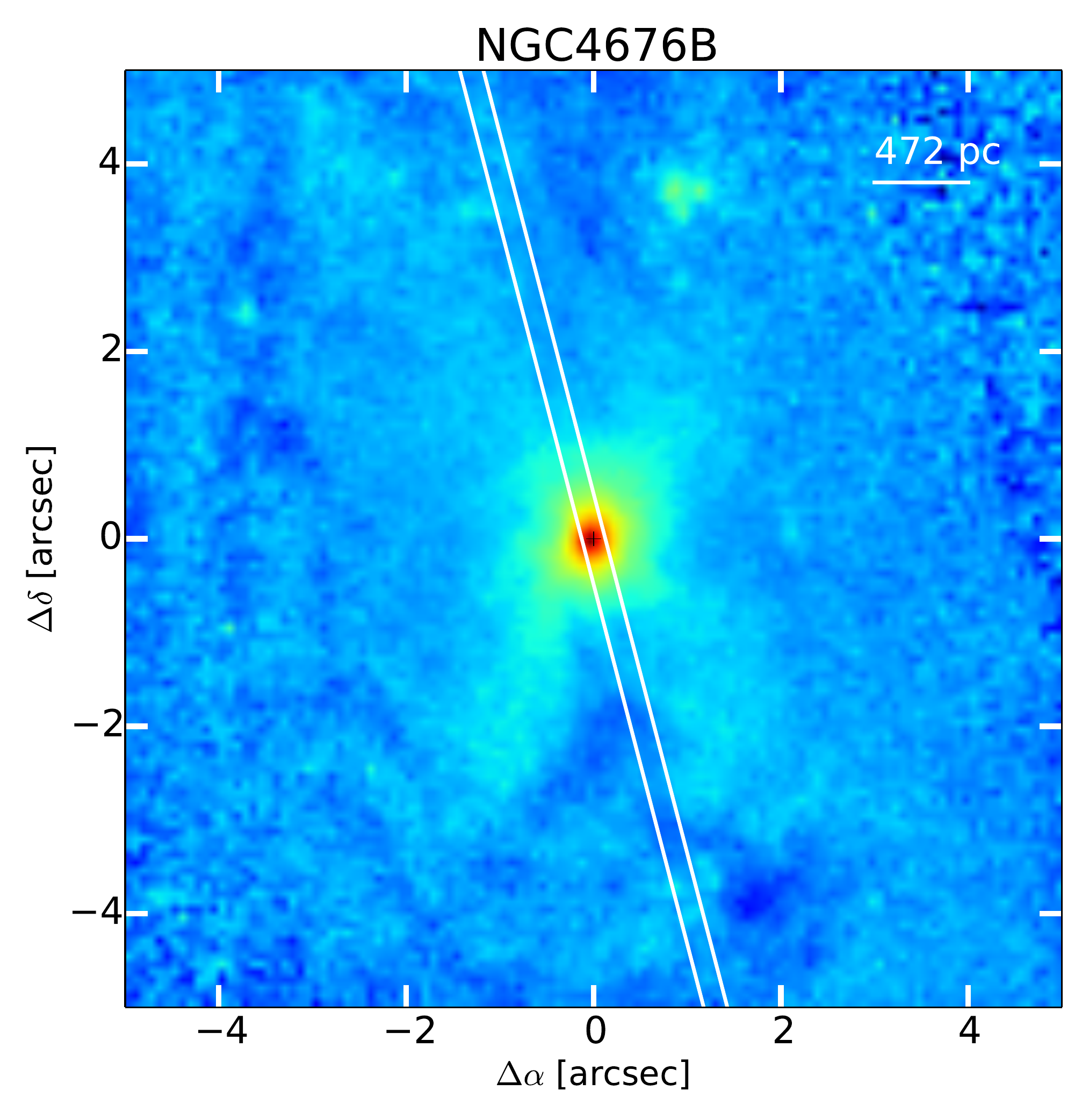}
   \includegraphics[width=\columnwidth]{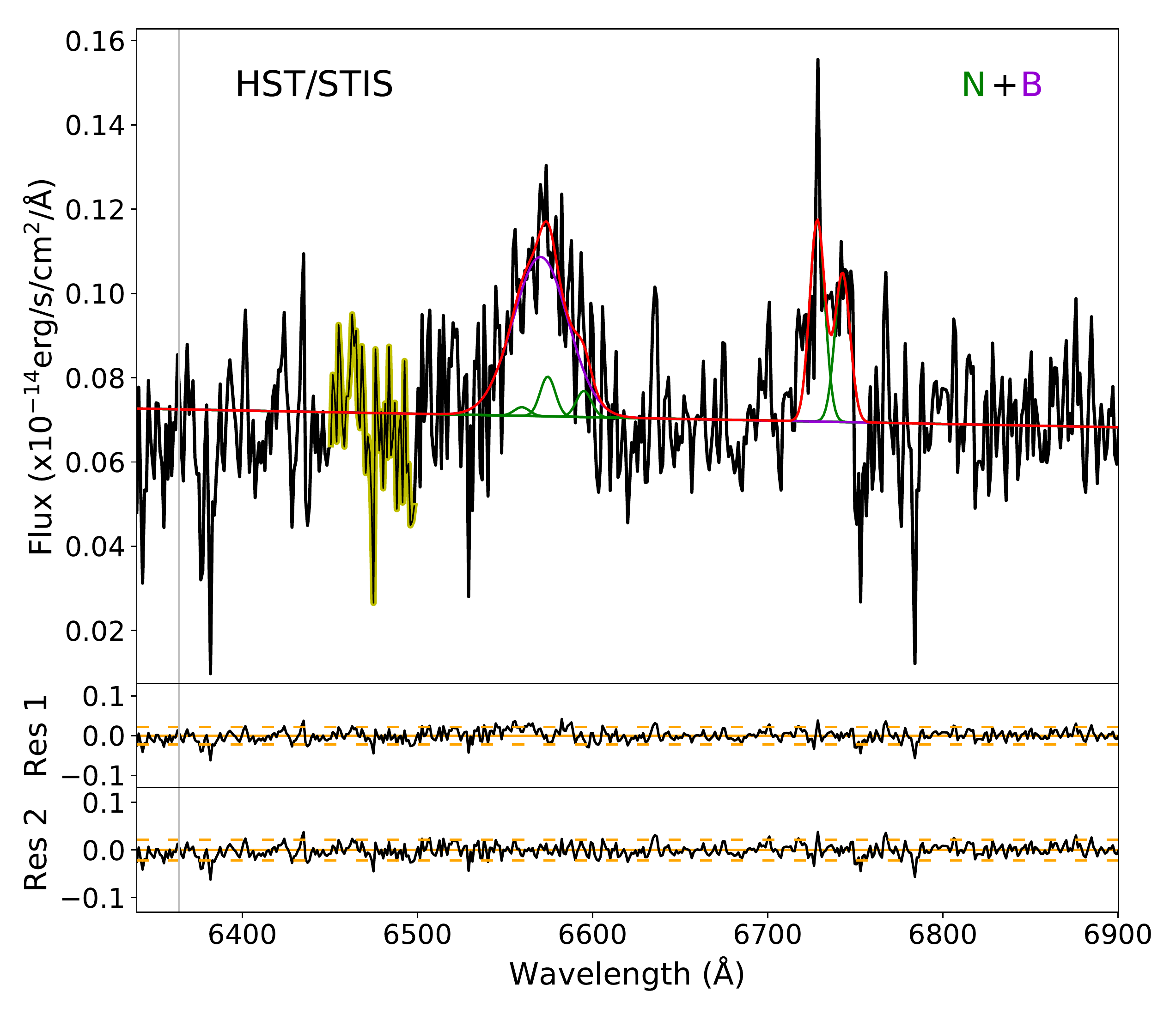}
 
 \caption{(General description in Appendix~\ref{appendix}).
         NGC 4676B: The S\,/\,N of the \textit{HST}/STIS spectrum is generally low and the quality of the fitting could be hampered. Both this and the wavelength range starting at 6340 \AA\, make the [O\,I] lines not visible, so only the S-method could be applied. The fit slightly improves by adding a broad component. There is no Palomar spectra for this galaxy. The PA is not indicated as this galaxy belongs to an interacting system and there is no clear determination of it.}
   \label{Panel_NGC4676B} 		 		 
\end{figure}

\begin{figure}
\centering
   \includegraphics[width=.4\textwidth]{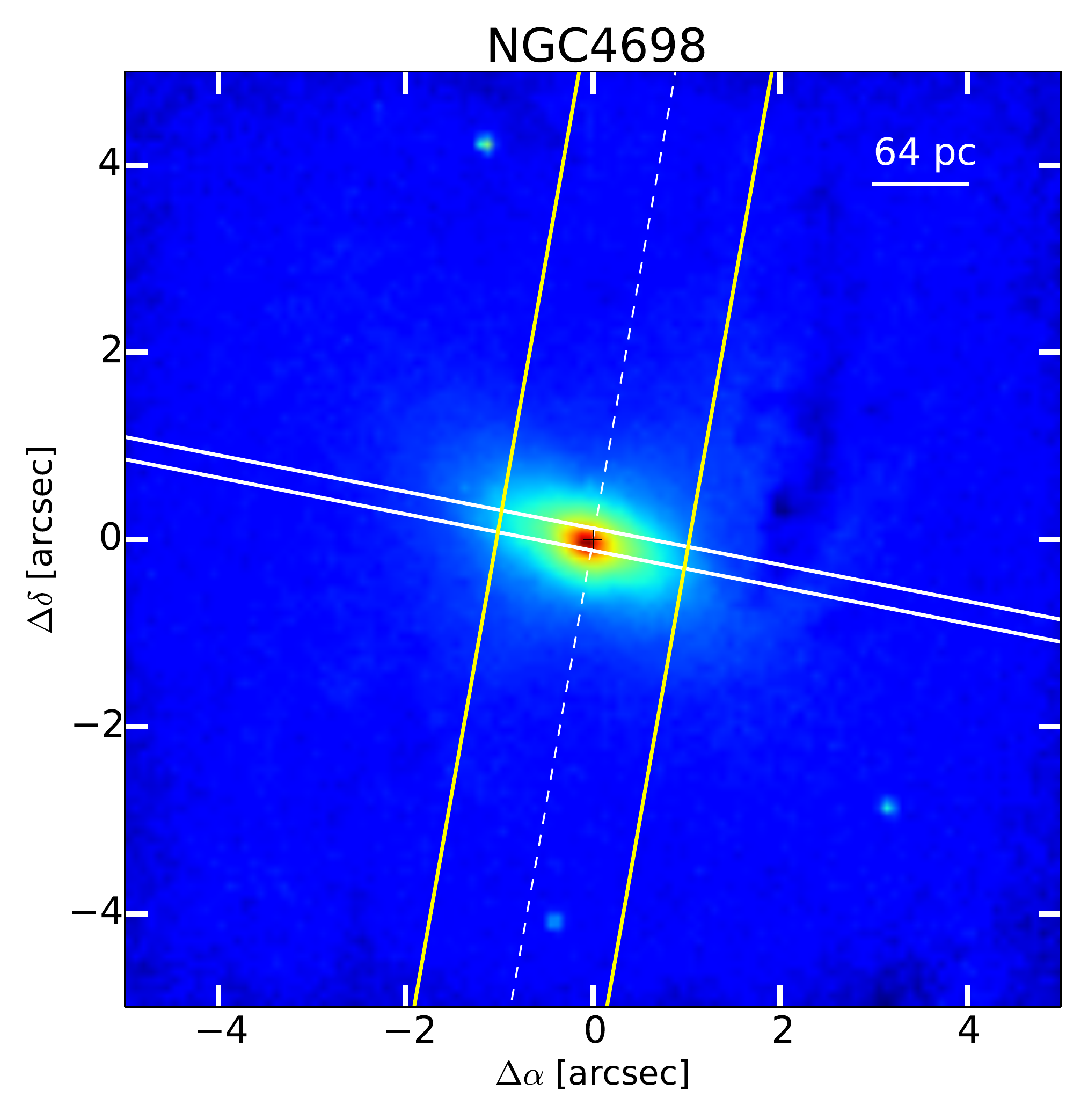}
   \includegraphics[width=\columnwidth]{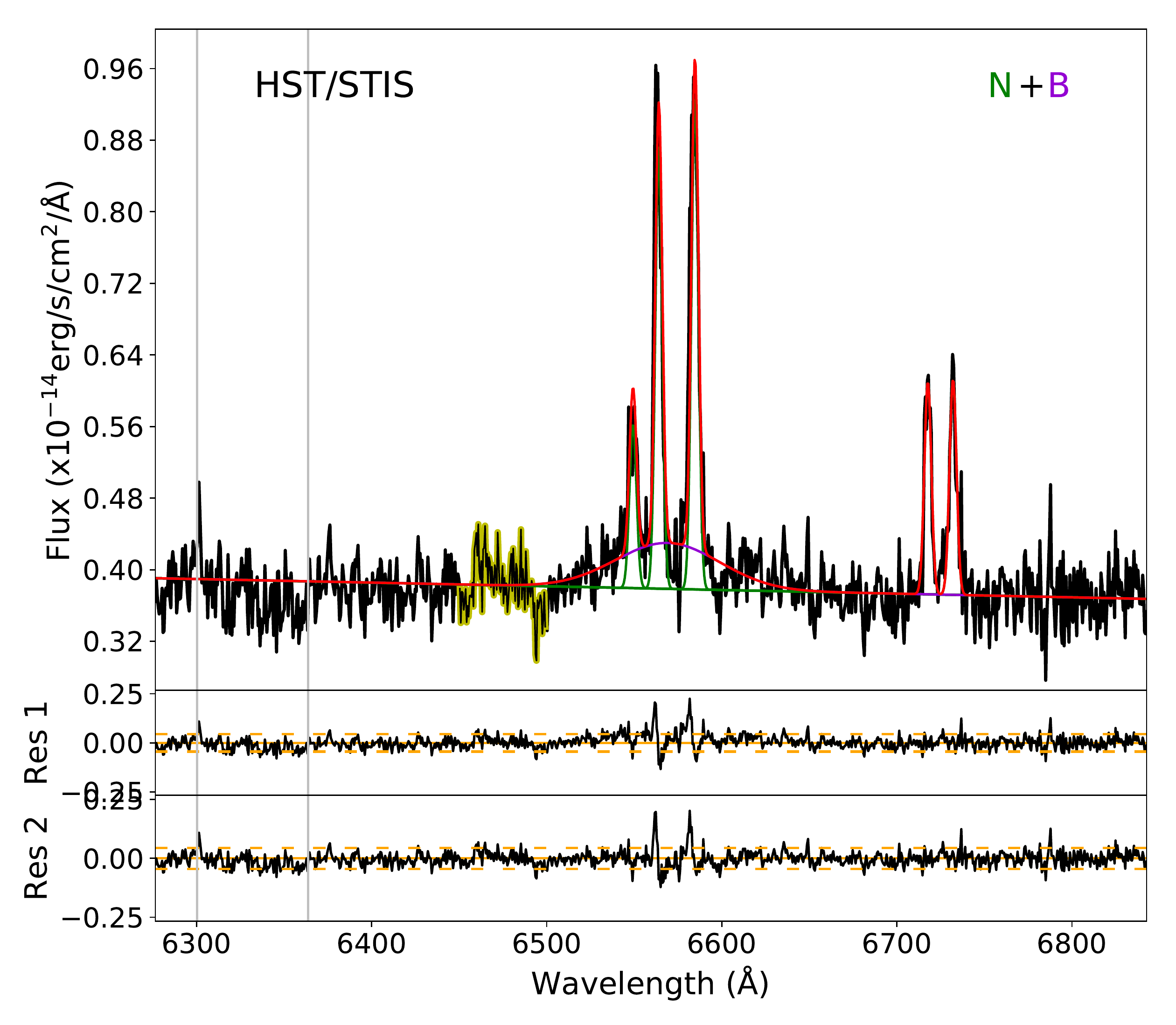}
   \includegraphics[width=\columnwidth]{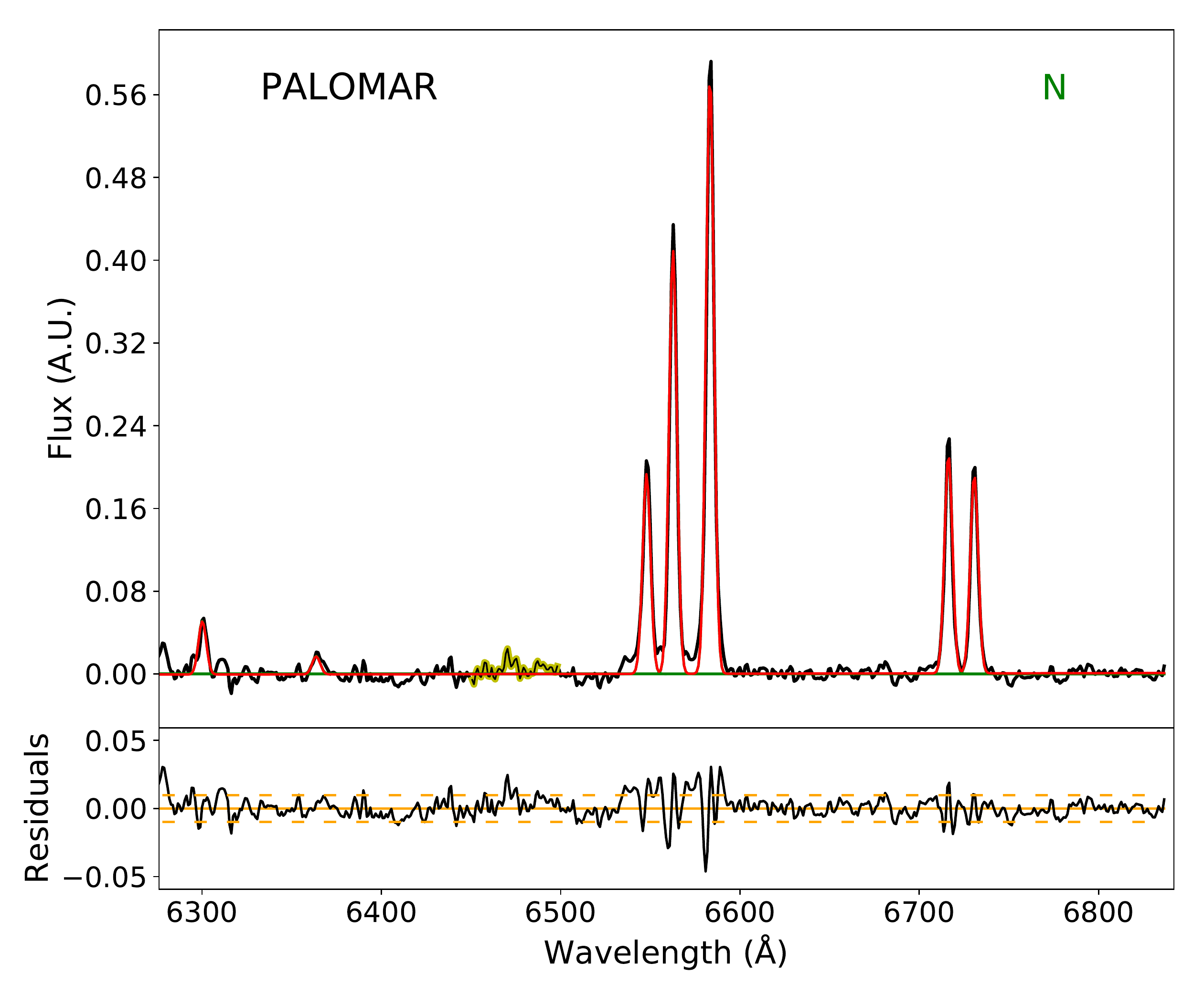}

 \caption{(General description in Appendix~\ref{appendix}). 
         NGC 4698: [O\,I] lines are not visible in the \textit{HST}/STIS spectrum. A narrow Gaussian component is sufficient to fit [S\,II] lines, and a broad component in H$\alpha$ to improve the modeling, although its contribution to the global fit is rather weak. For the Palomar data a single narrow component is sufficient to model the lines.}
   \label{Panel_NGC4698} 		 		 
\end{figure}

\begin{figure}
\centering
   \includegraphics[width=.4\textwidth]{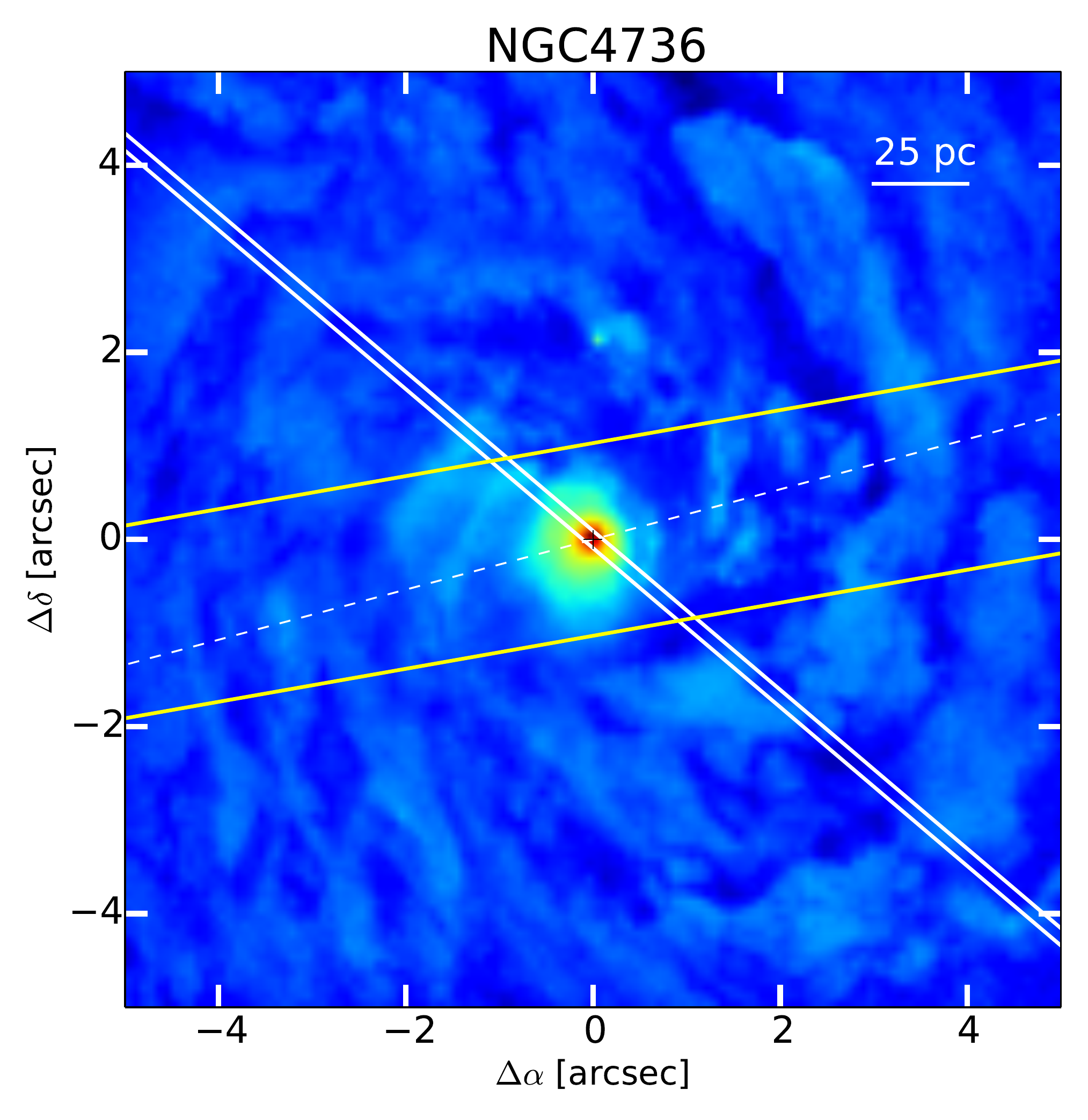}
   \includegraphics[width=\columnwidth]{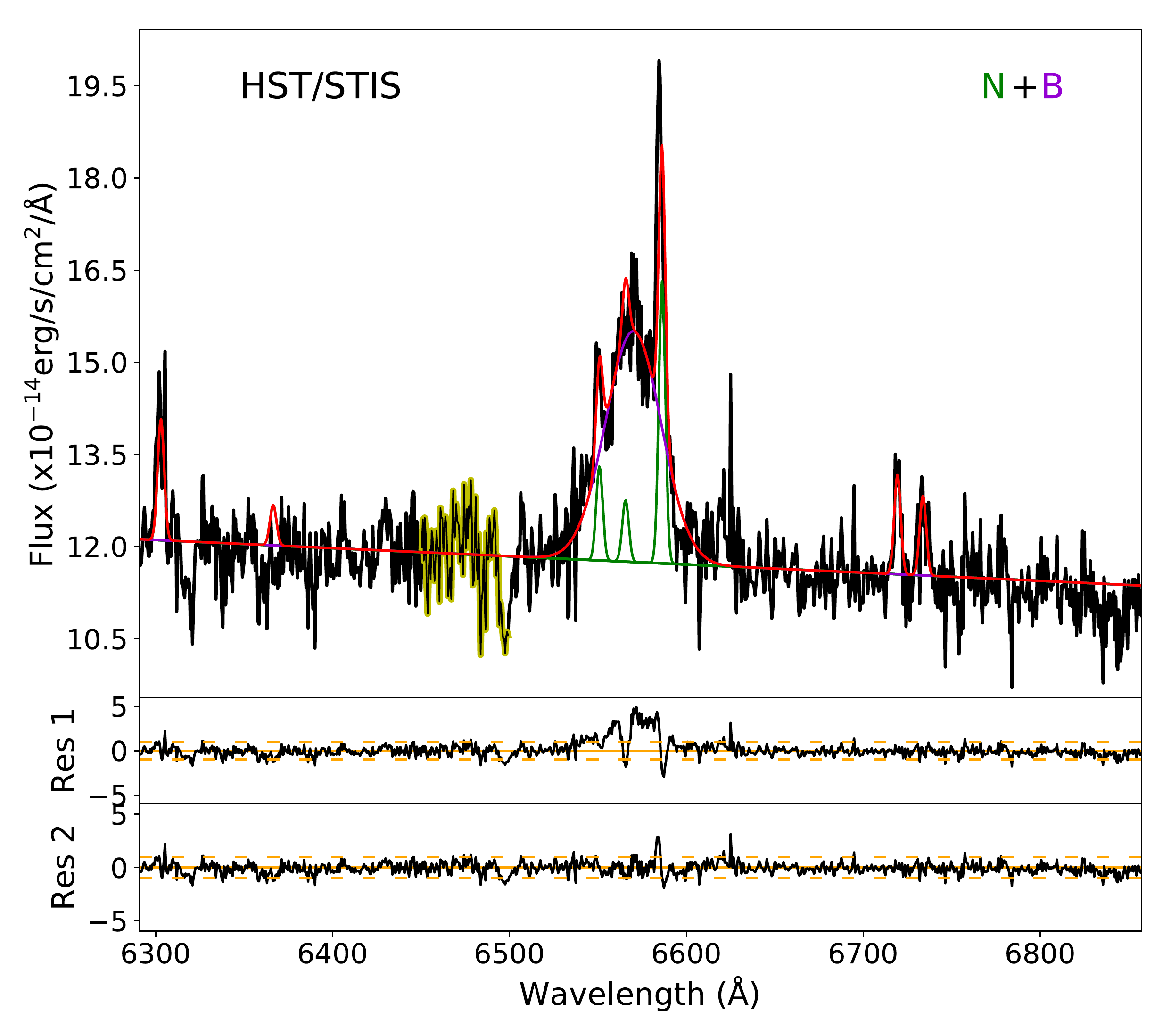} 
   \includegraphics[width=\columnwidth]{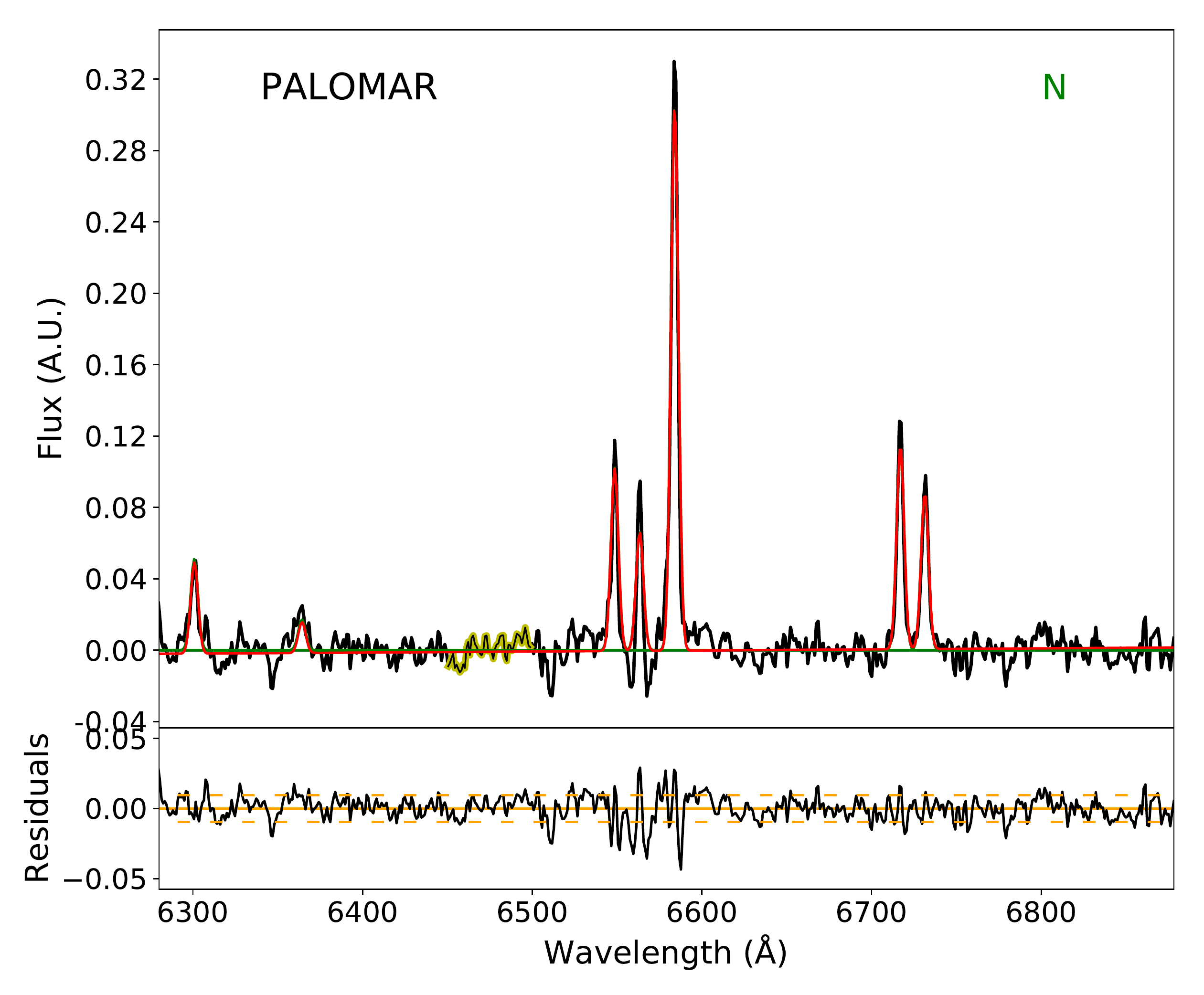}
 
 \caption{(General description in Appendix~\ref{appendix}). 
         NGC 4736: Both [S\,II] and [O\,I] lines are unblended in the \textit{HST}/STIS spectrum and they clearly present narrow profiles. In the H$\alpha$-[NII] lines a broad component is visible, as already reported by \citet{Constantin2015}. For the Palomar spectrum a narrow component is sufficient for fitting all the emission lines considered in our study. }
   \label{Panel_NGC4736} 		 		 
\end{figure}


\end{appendix}

\end{document}